\RequirePackage{fix-cm}
\documentclass[twocolumn,epjc3]{svjour3}  
\smartqed  
\RequirePackage{graphicx}
\RequirePackage{amsmath}
\RequirePackage{subfig}

\journalname{Eur. Phys. J. C}
\begin{document}

\title{Layouts for fixed-target experiments and dipole moment measurements of short-living baryons using bent crystals at the LHC}

\author{D. Mirarchi \thanksref{e1,addr1,addr2}
        \and  
        A. S. Fomin \thanksref{addr3, addr1} 
        \and  
        S. Redaelli \thanksref{addr1} 
        \and
        W. Scandale \thanksref{addr1}
}

\thankstext{e1}{e-mail: daniele.mirarchi@cern.ch}

\institute{CERN, European Organization for Nuclear Research, CH-1211 Geneva 23, Switzerland \label{addr1}
           \and
           currently also at The University of Manchester, Manchester M13 9PL, United Kingdom \label{addr2}
           \and
           NSC Kharkiv Institute of Physics and Technology, 1 Akademicheskaya St., 61108 Kharkiv, Ukrain \label{addr3}
           }

\date{Received: date / Accepted: date}

\maketitle


\begin{abstract}

Several studies are on-going at CERN in the framework of the Physics Beyond Collider study group, with main aim of broadening the physics research spectrum using the available accelerator complex and infrastructure. The possibility to design a layout that allows fixed-target experiments in the primary vacuum of the CERN Large Hadron Collider (LHC), without the need of a dedicated extraction line, is included among these studies. The principle of the layouts presented in this paper is to deflect beam halo protons on a fixed-target placed in the LHC primary vacuum, by means of bent crystals (i.e. crystal channeling). Moreover, interaction products emerging from the target can be used to perform electromagnetic dipole moments measurements of short-living baryons. Two possible layouts are reported, together with a thorough evaluation on their expected performance and impact on LHC operations.


\end{abstract}

\begin{figure*}[!ht]
\centering
\includegraphics*[width=100mm, height=35mm]{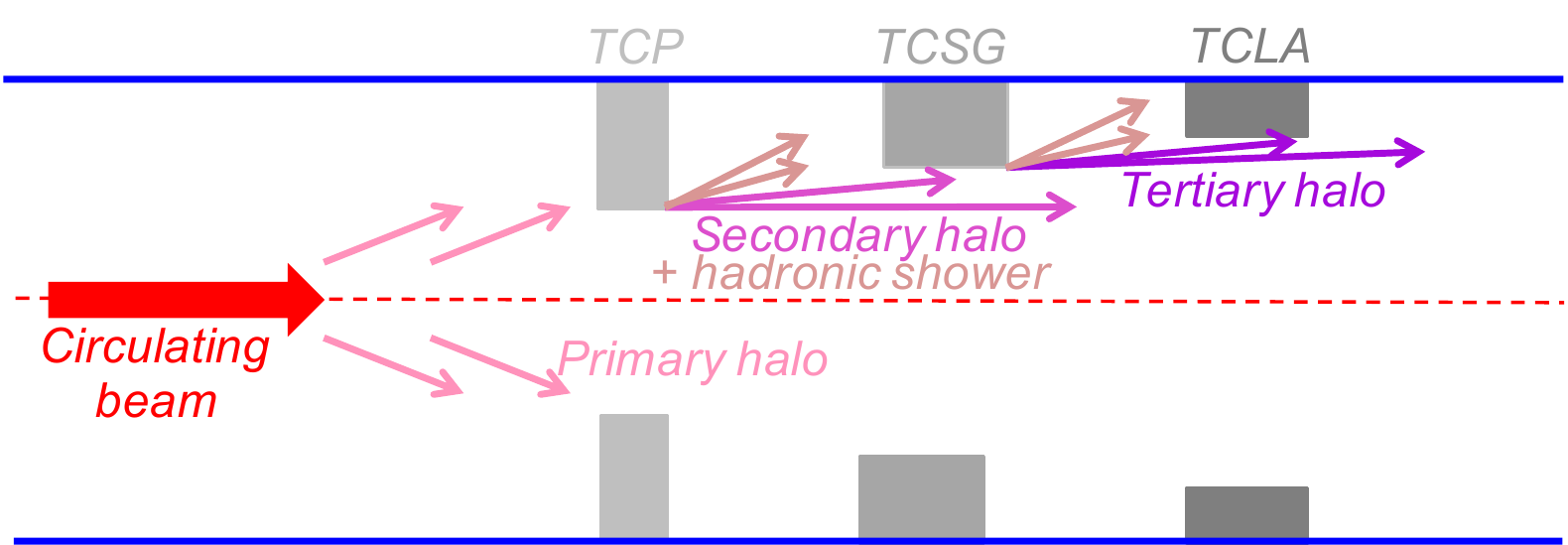}
\caption{Working principle of the current collimation system.}
\label{fig:std_coll}
\vspace*{-\baselineskip}
\end{figure*}

\begin{figure}[!ht]
   \centering
   \includegraphics[width=85mm]{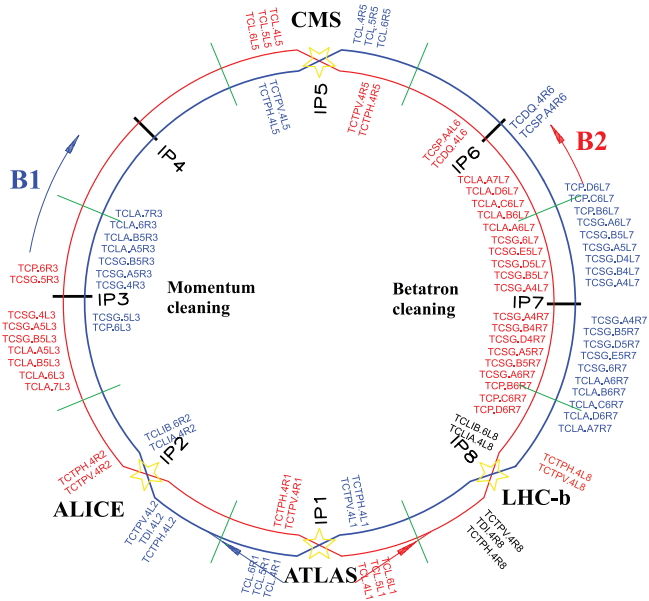}
   \caption{Collimation layout in the LHC, for both beams.}
   \label{fig:full_coll}
   \vspace*{-\baselineskip}
\end{figure}

\section{Introduction}

Several studies are on-going at CERN in the framework of the Physics Beyond Collider study group. The main aim is to assess the potential of the CERN accelerator complex and infrastructure to expand the physics reach beyond high-energy colliders. A powerful probe for studies of physics beyond the Standard Model is the measurement of electromagnetic dipole moment. Standard techniques consist of applying a dipolar magnetic field that induces a dipole-moment precession. The distribution of decay products depends on the induced precession. Thus, the dipole moment can be inferred by measuring such distribution. However, it is impossible to use conventional magnets for short-living baryons such as the $\mathrm{\Lambda_c}$, because the achievable magnetic field does not induce a measurable precession. A possible solution to overcome this problem is the use of bent crystals~\cite{baryshevsky1979spin,lyuboshits1979spin}. The equivalent magnetic field acting on a particle trapped between bent crystalline planes can be several orders of magnitude higher than what is achievable using dipole magnets, inducing measurable precession over distances of a few cm. This technique has been proved by the E761 Collaboration, which used the extracted 800~GeV/c proton beam from the Fermilab Proton Center on copper target to produce $\mathrm{\Sigma^+}$ and measuring its magnetic moment precession in bent crystals~\cite{chen1992first}.

A 6.5~TeV proton beam is nowadays available at the LHC, but no extraction lines are present. The experience gained with bent crystals for collimation of the circulating beam, triggered the idea of an in-vacuum fixed-target apparatus. Bent crystals can be used to deflect halo particles from the circulating beam onto a target placed in the LHC primary vacuum, allowing a unique opportunity for fixed-target experiments at such a high energy. The successful observation of crystal channeling with 6.5~TeV proton beams has been already achieved~\cite{scandale2016observation}. Heavier interaction products would become accessible at this energy, making possible to perform dipole-moment measurements of the $\mathrm{\Lambda_c}$. The main idea is to use a bent crystal to deflect halo particles of the circulating beam onto a target, where $\mathrm{\Lambda_c}$ are produced and channeled by a second bent crystal placed right after the target. This idea was firstly presented at the Physics Beyond Collider kickoff workshop~\cite{PBCkickoffWalter,PBCkickoffAchille} and gathered significant interest motivating further work on machine studies to conceive optimized layouts. Similar investigations were later carried out in~\cite{bagli2017electromagnetic}, also proposing to measure the electric-dipole moment of short-living baryons. The main scope of this paper is to assess the feasibility of this experiment from the accelerator physics side, comparing the expected performance of two possible layouts.

Very promising results towards a feasibility demonstration of the double-crystal concept were achieved by the UA9 collaboration at the CERN Super Proton Synchrotron (SPS), where a complete test stand has been setup~\cite{scandale2018ua9}. In 2017, the double channeling was demonstrated for the first time, by placing a crystal into the halo channeled by a first crystal in a setup equivalent to that proposed for the LHC experiment, although  still without target. In 2018, additional measurements were carried out in the SPS by adding to the setup a target upstream of the second crystal. 

\vspace{-0.3cm}

\section{LHC layout and collimation system}

\vspace{-0.2cm}

The LHC is a very complex and delicate machine that demands a tight control of beam loss because of its cryogenic nature. Tens of $\mathrm{mJ/cm^3}$ deposited in superconducting magnets can cause an abrupt loss of their superconducting properties, i.e. a magnet quench. On the other hand, about 300 MJ are presently stored in the LHC circulating beam, which will increase to about 700 MJ in the High-Luminosity upgrade (HL-LHC) \cite{apollinari2016high,TDRHL,PDRHL}. A highly-efficient collimation system is required in order to minimize the amount of deposited energy in the superconducting magnets by beam loss. 

\begin{figure*}[!ht]
\centering
\includegraphics*[width=160mm, height=43mm]{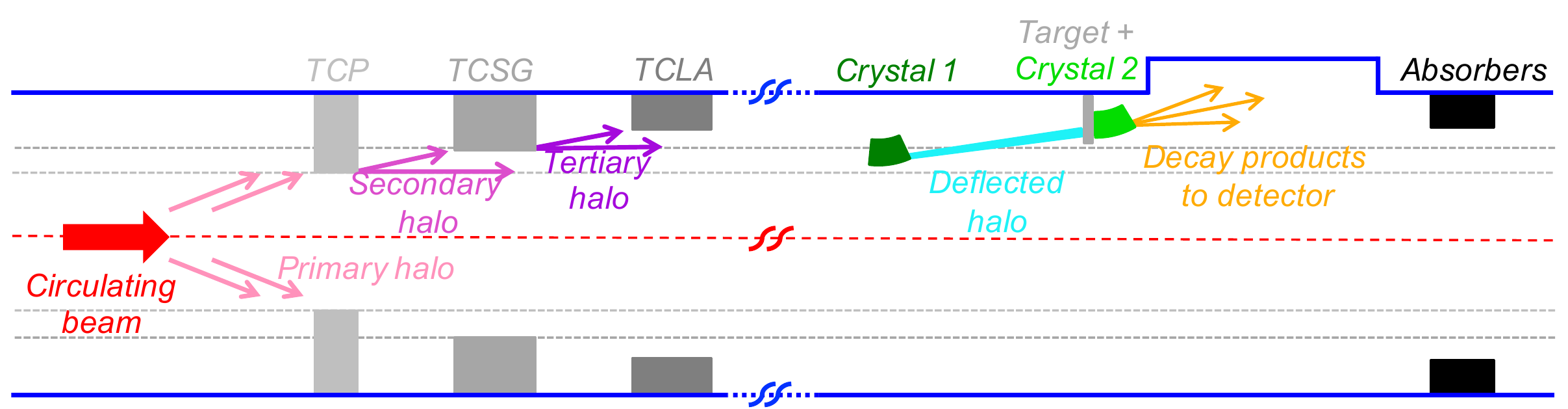}
\caption{Working principle of the double-crystal scheme for fixed-target experiments and dipole-moment measurements at the LHC, and its integration in the collimation hierarchy.}
\label{fig:dc_princ}
\vspace*{-\baselineskip}
\end{figure*}

An illustrative picture of the working principle of the LHC collimation system is given in Fig.~\ref{fig:std_coll}. The present LHC system \cite{des_coll} is composed of 44 movable ring collimators per beam, placed in a precise multi-stage hierarchy that must be maintained in any machine configuration to ensure optimal cleaning performance. Two LHC insertions (IRs) are dedicated to beam halo collimation: IR3 for momentum cleaning, i.e. removal of particles with a large energy offset (cut from $\delta p/p\sim0.2\:\%$ for zero betatron amplitude); and IR7 for betatron cleaning, i.e. continuous controlled disposal of transverse halo particles. Each collimation insertion features a three-stage cleaning based on primary collimators (TCP), secondary collimators (TCSG) and absorbers (TCLA). In this scheme, the energy carried by the beam halo intercepted by TCPs is distributed over several collimators (e.g. 19 collimators are installed in the betatron cleaning insertion). Dedicated collimators for protection of sensitive equipment (such as TCTP for the inner triplets), absorption of physics debris (TCL) and beam injection/dump protection (TDI/TCDQ-TCSP) are also present at specific locations of the ring. A detailed description of these functionalities goes beyond the scope of this paper and can be found in~\cite{des_coll}. The complete collimation layout of the LHC is shown in Fig.~\ref{fig:full_coll}.

The other IRs house the Radio Frequency system (RF) and the Beam Dump system (LBDS) in IR4 and IR6, respectively. The main physics detectors are placed in the remaining IRs: the multi-purposes ATLAS and CMS are placed in IR1 and IR5, respectively; the flavour physics LHCb is placed in IR8; the heavy ion physics ALICE is placed in IR2.

A natural choice to deploy a layout that would allow fixed-target and dipole-moment experiments is IR8, because of the presence of the LHCb detector that is suited for forward physics thanks to its asymmetric design. This option was originally studied in~\cite{PBCkickoffWalter}, where it was already clear that serious limitations could come from losses on superconducting magnets. In particular, a sort of mini collimation system would be needed around IR8 to handle losses of high-intensity beams. 

As opposed to the choice of installing new collimators around an existing large detector, new studies have been performed to probe the feasibility of using an existing collimation insertion where a smaller and dedicated detector could be located. In this case, the natural choice is IR3 for different reasons illustrated in section~\ref{sec:IR3}.


\section{Design goals, constraints and tools}
\label{sec:goal_cons}


The schematic working principle of the layouts is shown in Fig.~\ref{fig:dc_princ}. The main goal is to maximize the number of protons on target (PoT) while keeping the losses on superconducting magnets below limits tolerable for operations.

It would be possible to design a dedicated beam optics with present magnet layout, in order to optimize the layout performance. However, the overhead would become too large if new optics needs to be commissioned. It was therefore decided to design optimized layouts for the existing and already commissioned optics. The same approach has been used to design the crystal collimation layout currently installed in the LHC~\cite{mirarchi2017design}, which led to the successful observation of crystal channeling with 6.5~TeV protons beams in the LHC~\cite{scandale2016observation}.

\subsection{Main constraints}

Important constraints for the design, e.g. on longitudinal positions, come from space availability. Although this work does not include a dedicated integration study, all known constraints from the present space occupancy were taken into account in the layouts presented here.


Regarding the best location of the first crystal, once free locations are found they must be combined with optimal beam optics requirements in order to: 

\begin{itemize}

\item Enhance the displacement due to a given deflection by maximizing the beam size (i.e. $\sigma(s)= \sqrt{\beta(s) \varepsilon}$).

\item Improve the channeling efficiency of multi-turn halo by minimizing the beam divergence (i.e. $\sigma^\prime(s)= \sqrt{\gamma(s) \varepsilon}$).

\end{itemize}

\noindent Thus, the following ratio must be maximized:

\begin{equation}
\frac{\sigma(s)}{\sigma^\prime(s)}=\sqrt{\frac{1}{1+\alpha^2(s)}}\:,
\end{equation}

\noindent where $\alpha$,  $\beta$ and $\gamma$ are the Twiss parameters, while $\varepsilon$ is the physical beam emittance. This ratio is 0.91 for both layouts presented in sections \ref{sec:IR8} and \ref{sec:IR3}, which is very important to enhance the channeling efficiency of secondary and tertiary halo, i.e. halo protons emerging from primary and secondary collimators, respectively (see Fig.~\ref{fig:dc_princ}). 


It is clear that absorbers need to be added to dispose of the channeled halo that emerges from the target and to dispose of out-scattered protons. Optimizing the location for such collimators calls for an installation at the closest location where the betatron phase advance from the first crystal is about $\pi/2$. Smaller beam sizes at the collimators are also favoured because allow closer settings for a better efficiency in intercepting particles emerging from crystal and target. These two parameters -- phase difference and transverse settings -- define the angular cut made by the absorber, as defined in Eq. (\ref{eq:kick-traj_LHC}). For convenience, this is typically expressed as the minimum kick for particles out-scattered at the crystal that are still intercepted by the downstream collimators. The optimization of this angular cut is one of the main differences between the layouts presented in sections \ref{sec:IR8} and \ref{sec:IR3}.

Between the first crystal and the absorber, it is required that the trajectory of channeled halo particles remains at least 4~mm apart from the geometrical LHC aperture, following the guidelines from the LHC Technical Design Report~\cite{bruning2004lhc}. Detailed aperture calculations, with a proper accounting of relevant errors on optics, aperture, orbit, etc., shall be performed in a later stage. A distance between the target and the circulating beam envelope of 4~mm is also required, which defines the minimum bending angle of the first crystal. This retraction ensures that the target is not intercepting significant beam halo. It can be demonstrated that, for the operational scenarios discussed below, approaching the target further does not bring significant benefits.




%
%
%
%
%

Furthermore, all the system must be placed in the vertical plane in order to relax constraints due to machine protection aspects. This because the beam is deflected in the horizontal plane and directed into the dump line when a beam dump is triggered. It can happen that the beam dump kickers are not fired synchronously with respect to the abort gap (range of the ring left empty) and dangerous portions of the beam are kicked wrongly. Thus, by placing our system in the vertical plane the possibility to get hit by the beam during an asynchronous dump is removed. This opens the possibility to get closer to the beam with the first crystal, in principle down to the aperture of the primary collimators in IR7. Finally, crystals must be placed on the top side in order to fit the goniometers needed to place and orient them\cite{LHC_gonio2,LHC_gonio3}.

\begin{table*}
\centering
	\caption{Installation position and main features of the proposed experimental layout in IR8. All the components act on the vertical plane.}
	\label{IR8_summ}
		\begin{tabular}{cccccc}
		\hline\noalign{\smallskip}
		\textbf{Name}  & \textbf{$s$  from IP1}     & \textbf{Bending} & \textbf{Length} & \textbf{Mat.} & \textbf{Bending}\\
	 &	 \textbf{[m]} & \textbf{[$\mu$rad]} & \textbf{[cm]} & &\textbf{planes} \\
		\noalign{\smallskip}\hline\noalign{\smallskip}
		 $\mathrm{Cry_1}$ & 23220  & 150 & 1.2 & Si & 110 \\
	Target & 23313  & - & 0.5 & W & - \\
	$\mathrm{Cry_2}$ & 23313  & 14000 & 7 & Si & 110 \\
	TCSG.A4R8.B1 & 23402 &  - & 100 & CFC & - \\
	TCSG.B4R8.B1 & 23404 &  - & 100 & CFC & - \\
	TCSG.C4R8.B1 & 23406 &  - & 100 & CFC & - \\
	TCLA.A4R8.B1 & 23408 &  - & 100 & W & - \\
\noalign{\smallskip}\hline
\end{tabular}
\vspace{-\baselineskip}
\end{table*}

\subsection{Simulation tools}
\label{sec:sim_tool}

Semi-analytical tools were developed to evaluate quickly the feasibility of each layout. The trajectory of particles experiencing an angular kick ($\theta$) at $s_{1}$ can be described using the transfer matrix formalism. If a crystal is installed at $s_{1}$, and assuming $\alpha_{1}\sim0$, the trajectory of a kicked particle is described by:

\begin{align} \label{eq:kick-traj_LHC}
x(s)&=\sqrt{\frac{\beta(s)}{\beta(s_{1})}}\cos\left(\Delta\mu_{s-s_{1}}\right)x(s_{1})\\
&+\theta\sqrt{\beta(s)\beta(s_{1})}\sin\left(\Delta\mu_{s-s_{1}}\right)\:, \nonumber
\end{align}

\noindent where $\Delta\mu$ is the phase advance and other parameters were defined above. 

After a first identification of suitable installation locations based on space availability, a sub-set of possible locations, crystal parameters, and collimator settings is determined based on semi-analytical tools. Then, complete multi-turn tracking simulations are performed. They are made using \texttt{SixTrack} \cite{SixT_man,SixT_man2,Coll_SixT_1,Coll_SixT_2} that allows a symplectic, fully chromatic and 6D tracking along the magnetic lattice of the LHC, taking into account interactions with the ring collimators and the detailed aperture model of the entire machine. The treatment of interactions between protons and bent crystals is carried out using a dedicated routine \cite{Vale,mirarchi2015crystal,mirarchi2015crystal_rout,routineYellow} implemented in  \texttt{SixTrack}. This simulation setup allows estimation of the density of protons lost per metre with a resolution of 10 cm along the entire ring circumference, for a given halo intercepted and collimator settings.

\begin{figure}[t]
   \centering
   \includegraphics[width=90mm]{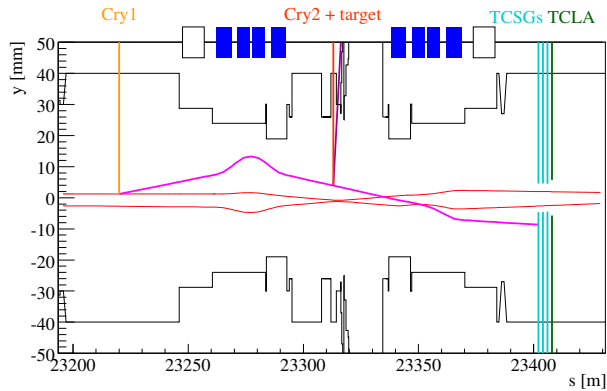}
   \caption{Trajectory of channeled halo particles (light magenta line) and mechanical aperture of the beam pipe (black) versus longitudinal position along the IR8 insertion. The first crystal, shown by the light orange line, sits on the $5~\sigma$ beam envelope shown by the red lines. The assembly of target plus second crystal, shown by the dark orange line, is placed to intercept channeled halo particles and the trajectory followed by particles channeled is shown by the dark magenta line. The TCSGs and TCLA used to intercept channeled halo particles are reported in cyan and green lines and are set at $10~\sigma$ and $13~\sigma$, respectively. The magnetic lattice is also reported on top, where blue and white boxes represent main superconducting quadrupoles and dipoles, respectively.}
   \label{fig:IR8_lay}
\end{figure}

\section{IR8 layout design}
\label{sec:IR8}

The design of optimized layout in IR8 is shown in Fig.~\ref{fig:IR8_lay}, which is placed on Beam 1 with clock-wise orientation due to the LHCb asymmetry. It consists of a first crystal ($\mathrm{\mathrm{Cry_1}}$) with bending $\theta_b^{\mathrm{\mathrm{Cry_1}}}=150~\mu$rad that separates the halo particles from the primary beam sufficiently to impinge on a target, respecting the constrains of displacement of 4 mm with respect to the beam envelope at the target, and clearance between the geometrical machine aperture and the deflected halo of at last 4 mm along the entire trajectory.

The crystal length is $l^{\mathrm{Cry_1}}=1.2~$cm, which was chosen to have a bending radius $R=80~$m, as the present crystal installed in the LHC. This follows the parametric studies reported in~\cite{mirarchi2017design} and ensures an optimum crystal channeling performance at LHC top energy, while keeping the nuclear interactions rate as low as possible. 

A target is placed upstream of the LHCb detector at about 2.4~m from IP8, which is the closest free slot where about 1.3~m of longitudinal space are available to fit a tank. It is important to be as close as possible to the Vertex Locator (VELO) to obtain the best resolution on decay vertex. 

The optimal target length and material should be tuned to fulfill the physics requirements. A 5 mm long target of tungsten is presently assumed for the production of $\mathrm{\Lambda_c}$, in order to increase its production cross section while keeping low detrimental effects (i.e. decay and multiple coulomb scattering in the target volume)~\cite{fomin2017feasibility,samsonov1996possibility}.

A second crystal ($\mathrm{Cry_2}$) is placed adjacent to the target. The required bending is $\theta_b^{\mathrm{Cry_2}}=14~$mrad, which is needed to send $\mathrm{\Lambda_c}$ decay products inside the LHCb acceptance, while the length is $l^{\mathrm{Cry_2}}=7~$cm, as defined in~\cite{bagli2017electromagnetic}. Both $\mathrm{Cry_1}$ and $\mathrm{Cry_2}$ are presently considered to be made of silicon, because it is produced with high purity lattice and dislocations below $1/\mathrm{cm^{2}}$. germanium crystals can feature a similar lattice quality and a deeper potential well, which could slightly improve the single-pass channeling efficiency. However, R\&D on bent silicon crystals is much more advanced and they are presently used in the LHC. Beam halo particles that do not interact with the target+$\mathrm{Cry_2}$ assembly are intercepted by 4 double-sided LHC-type collimators. The first 3 are made of 1~m long carbon-fiber-carbon composite jaws (as the present TCSGs in the LHC), while the last one is made of 1~m long tungsten jaws (as the present TCLAs in the LHC). Nevertheless, this is a performance-oriented choice whose feasibility needs to be discussed later and the number of absorbers needed can be revised as a function of the operational scenario. Parametric studies have been performed by changing the number and material of these absorbers and this configuration has been found to have the best performance with a minimal number of collimators. Their longitudinal position has been defined to optimize the angular cut on protons out-scattered by  $\mathrm{Cry_1}$. Protons that acquire an angular deflection $>60~\mu$rad by $\mathrm{Cry_1}$ are intercepted, if TCSGs are set at 10~$\sigma$ (minimum setting allowed to respect collimation hierarchy).

It is noted that this setup is only on one side of the beam, as initially proposed in~\cite{PBCkickoffWalter} and opposed to what proposed in~\cite{bagli2017electromagnetic}. This choice has no critical impact on the studies described here, and simplifies operational aspects and the overall complexity of the apparatus. Taking into account the multi-turn dynamics, our study indicates that the improvement from doubling the devices by installing them at both sides is minor. If $\mathrm{Cry_1}$ were to be used as primary collimator, it would intercept all particles diffusing out of the core so a second, symmetric apparatus would be useless. For retracted $\mathrm{Cry_1}$ settings the impact was minor for the optics of 2018 studied here \footnote{The motivation to use a double apparatus in the E761 experiment was to verify the result of measurement and to increase the statistics, but not to reduce the systematic uncertainty of measurement~\cite{Chen1992ai,Kou2019}. }.

The main layout parameters are reported in Table~\ref{IR8_summ}.


\section{IR3 layout design}
\label{sec:IR3}

\begin{figure}[t]
   \centering
   \includegraphics[width=90mm]{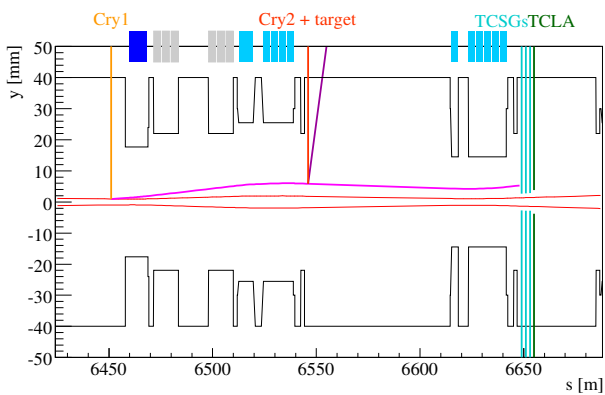}
   \caption{Proposed experimental layout in IR3. Same notation as in Fig.~\ref{fig:IR8_lay}. Warm quadrupoles and dipoles are shown by the light blue and gray boxes, respectively.}
   \label{fig:IR3_lay}
   \vspace*{-\baselineskip}
\end{figure}

\begin{table*}[ht]
\centering
	\caption{Installation position and main features of the proposed experimental layout in IR3. All the components act on the vertical plane.}
	\label{IR3_summ}
		\begin{tabular}{cccccc}
		\hline\noalign{\smallskip}
		\textbf{Name}  & \textbf{$s$ from IP1}    & \textbf{Bending} & \textbf{Length} & \textbf{Mat.} & \textbf{Bending}\\
	 &	 \textbf{[m]} & \textbf{[$\mu$rad]} & \textbf{[cm]} & &\textbf{planes} \\
		\noalign{\smallskip}\hline\noalign{\smallskip}
		 $\mathrm{Cry_1}$ & 6451  & 50 & 0.4 & Si & 110 \\
	Target & 6546  & - & 0.5 & W & - \\
	$\mathrm{Cry_2}$ & 6546  & 5000 & 7.5 & Si & 110 \\
	TCSG.A4R8.B1 & 6649  & - & 100 & CFC & - \\
	TCSG.B4R8.B1 & 6651  & - & 100 & CFC & - \\
	TCSG.C4R8.B1 & 6653  & - & 100 & CFC & - \\
	TCLA.A4R8.B1 & 6655  & - & 100 & W & - \\
\noalign{\smallskip}\hline
\end{tabular}
\end{table*}

The design of the optimized layout in IR3 is shown in Fig.~\ref{fig:IR3_lay}. The main motivation to look for an alternative layout comes from the need to overcome intrinsic limitations that are present in IR8, namely:

\begin{itemize}

\item Required bending of $\mathrm{Cry_1}$ and angular cut performed by the absorbers, due to the local optics in proximity of IP8.

\item Required bending of $\mathrm{Cry_2}$ to send decay products into the LHCb acceptance.

\end{itemize}

The IR3 layout consists of a $\mathrm{Cry_1}$ with bending $\theta_b^{\mathrm{Cry_1}}=50~\mu$rad and length $l^{\mathrm{Cry_1}}=4~$mm (same parameters are used for crystals presently installed in IR7 for collimation studies\cite{scandale2016observation,mirarchi2017design}). With respect to the layout in IR8, these parameters allow to reduce the nuclear interaction rate at the first crystal making accessible smaller settings that lead to a larger rate of PoT achievable. 

The same criteria as in IR8 has been used to design the absorber layout, which is made of the same elements. The reduced constraints on longitudinal space available and the smaller $\beta_y(s)$ function with respect to IR8, made possible to better optimize their performance. Using the same TCSGs settings of 10~$\sigma$ the angular cut performed is of about $20~\mu$rad (i.e. $\times3$ smaller than in IR8). This is very important in terms of operational performance, as discussed in section~\ref{sec:perf}. 

The bending of the second crystal can be significantly reduced, increasing the yield of channeled $\mathrm{\Lambda_c}$  that acquired the desired precession, as further discussed in section~\ref{sec:sec_cry}.

Same considerations as in IR8 apply to the target choice, i.e. 5~mm long tungsten in case of $\mathrm{\Lambda_c}$ studies. 

A drift space of about 70~m is present between target+$\mathrm{Cry_2}$ assembly and the first downstream magnet. Thus, a dedicated experimental apparatus could be designed and fit in this available space (with the reconstruction of the $\mathrm{\Lambda_c}$ decay products as main functionality needed).

Same considerations as in IR8 apply on the possibility of a mirrored $\mathrm{Cry_1}$+target+$\mathrm{Cry_2}$ assembly. 

Another important difference with respect to the IR8 layout is that the magnets in this insertion are warm, as opposed to the superconducting magnets in IR8. Thus, reduced constraints on sustainable beam loss and relative magnet lifetime are present in IR3. 

A clock-wise orientation in Beam 1 of the layout is adopted because possible debris from the absorbers goes to IR4 (where the RF is placed) rather than to IR2 (where the ALICE experiment is located). 

The main layout parameters are reported in Table~\ref{IR3_summ}.

An additional feature of IR3 is that the two beams are in two separated vacuum pipes and do not interfere with each other. Thus, one could consider to have a mirror layout in Beam 2 that shares a common detector in the 70~m drift space. This will require duplicating the hardware of crystals, target and absorbers but can open the possibility to either perform different studies in the two beams\footnote{e.g. target+$\mathrm{Cry_2}$ assembly optimized for electromagnetic moment measurement of different particles.} or double the statistics.

\vspace*{-2mm}

\section{Machine losses and achievable PoT}
\label{sec:perf}

\begin{table*}[ht]
\centering
	\caption{LHC operational parameters in 2018 at End of Squeeze.}
	\label{tab:opt_par}
	\begin{tabular}{ccccc}
	\hline\noalign{\smallskip}
	\textbf{IP}  & \textbf{$\beta^*$ [cm]}  & \textbf{Crossing angle [$\mu$rad] (plane)} & \textbf{Separation [mm] (plane)} & \textbf{IP displacement [mm] (plane)} \\
	\noalign{\smallskip}\hline\noalign{\smallskip}
	1 & 30 & 160 (V) & -0.55 (H) & 0\\
	2 & 1000 & 200 (V) & 1 (H) & -2 (V)\\
	5 & 30 & 160 (H) & 0.55 (V) & -1.8 (V)\\
	8 & 300 & -250 (H) & -1 (V) & 0\\
\noalign{\smallskip}\hline
\end{tabular}
\vspace*{-\baselineskip}
\end{table*}

\begin{table}[t]
   \centering
   \caption{LHC collimation settings in 2018.}
   \begin{tabular}{lcc}
       \hline\noalign{\smallskip}
       \textbf{Coll. Family} & \textbf{IR}                      & \textbf{Settings [$\sigma$]} \\
       \noalign{\smallskip}\hline\noalign{\smallskip}
           TCP/TCSG/TCLA         & 7            & 5.0 / 6.5 / 10       \\ 
           TCP/TCSG/TCLA         & 3            & 15 / 18 / 20	 \\ 
           TCTP	& 1 / 2 / 5 / 8		& 8.5 / 37 / 8.5 / 15	\\
           TCL	& 1 / 5		& OUT		 \\	
           TCSP/TCDQ	& 6 		& 7.4 / 7.4\\
       \noalign{\smallskip}\hline
   \end{tabular}
   \label{tab:collsettings}
   \vspace*{-\baselineskip}
\end{table}	

An extensive simulation study was carried out to assess comparatively the expected performance of the layouts proposed. Simulations were performed using the tools introduced in section \ref{sec:sim_tool}. The main goal is the evaluation of the loss pattern around the entire LHC ring, to be compared to the present operational configuration. This is very important in order to define a possible operational scenario. In particular, if the loss pattern is not affected by the insertion of the $\mathrm{Cry_1}$+target assembly, it would be possible to perform measurements during standard physics operations. This operational mode is defined as \emph{parasitic}. On the other hand, if significant amount of losses are induced by the presence of these objects, the maximum loss rate (i.e. maximum stored intensity) that ensures safe and reliable machine operations must be estimated. This operational mode is defined as \emph{dedicated}.

\vspace{-2mm}

\subsection{Machine configuration}

The machine configuration ``End of Squeeze'' is used for the simulations reported here. At this point of the LHC cycle the optics of the machine is the same as in physics (with colliding beams), but the separation bumps are not yet collapsed. Thus, it is the most critical configuration in terms of available geometrical aperture. The 2018 operational optics and 6.5~TeV proton beams have been used, with main parameters reported in Table~\ref{tab:opt_par}.

Operational settings for the entire collimation system were used and reported in Table~\ref{tab:collsettings}.


\subsection{Operational performance}

\begin{figure}[!t]
\subfloat[]{\includegraphics*[width=90mm]{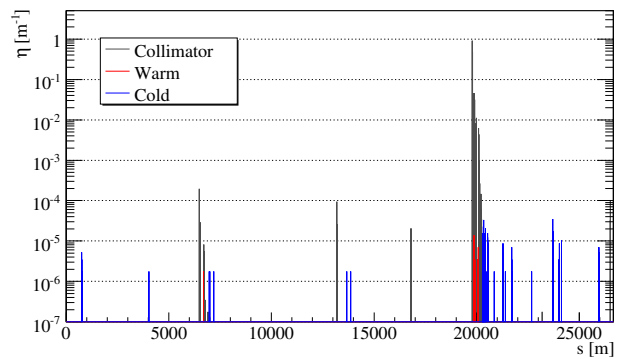}\label{fig:pres_coll_tot}}\vspace*{-\baselineskip}\\
\vspace*{-\baselineskip}
\subfloat[]{\includegraphics*[width=90mm]{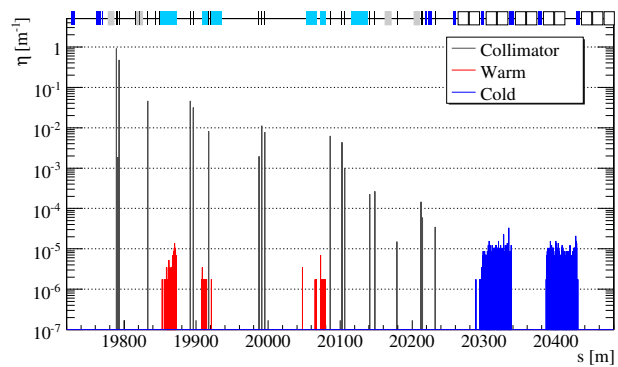}\label{fig:pres_coll_IR7}}
\caption{Simulated beam-loss pattern at End of Squeeze, with 6.5~TeV beams in 2018 and operational settings. The whole LHC (a), zoom of the IR7 insertion (b) (1~p~=~1.8$\times10^{-6}~\mathrm{m^{-1}}$).
}
\label{fig:pres_coll}
\vspace*{-\baselineskip}
\end{figure}

An example of simulated loss maps with 2018 operational settings is shown in Fig.~\ref{fig:pres_coll}. Losses on superconducting magnets, warm elements and collimators are indicated as cold, warm and collimator, respectively. It is clearly visible that the Dispersion Suppressor in IR7 (IR7-DS) is the limiting location of the whole ring in terms of cleaning efficiency (i.e. where the highest losses on cold elements are present). Thus, \emph{parasitic} operations can be envisaged if losses on cold magnets stays below present IR7-DS level after the insertion of the $\mathrm{Cry_1}$+target assembly.

\begin{figure}[!t]
\subfloat[]{\includegraphics*[width=90mm]{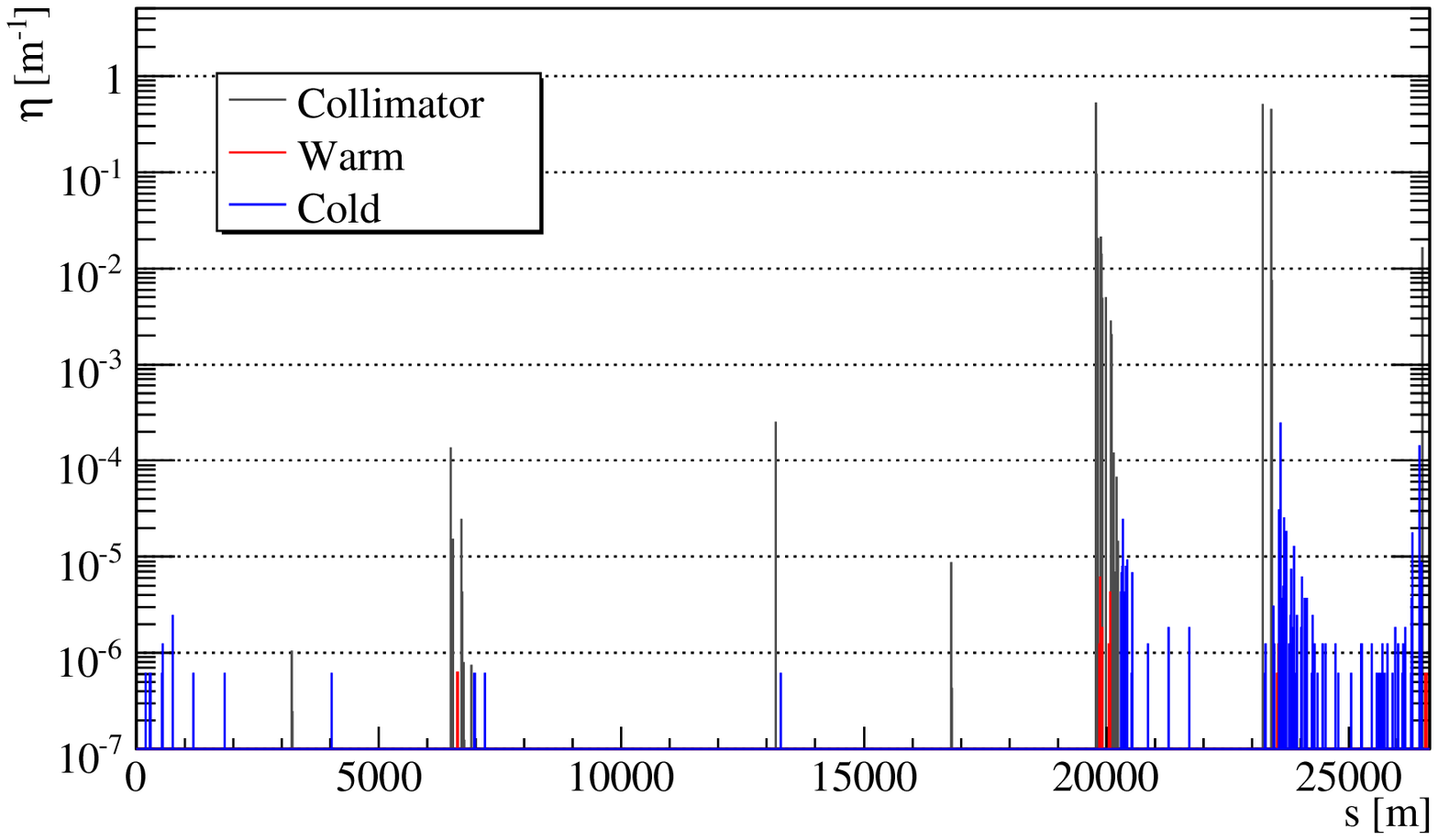}\label{fig:IR8_5s_tot}}\vspace*{-\baselineskip}\\
\vspace*{-\baselineskip}
\subfloat[]{\includegraphics*[width=90mm]{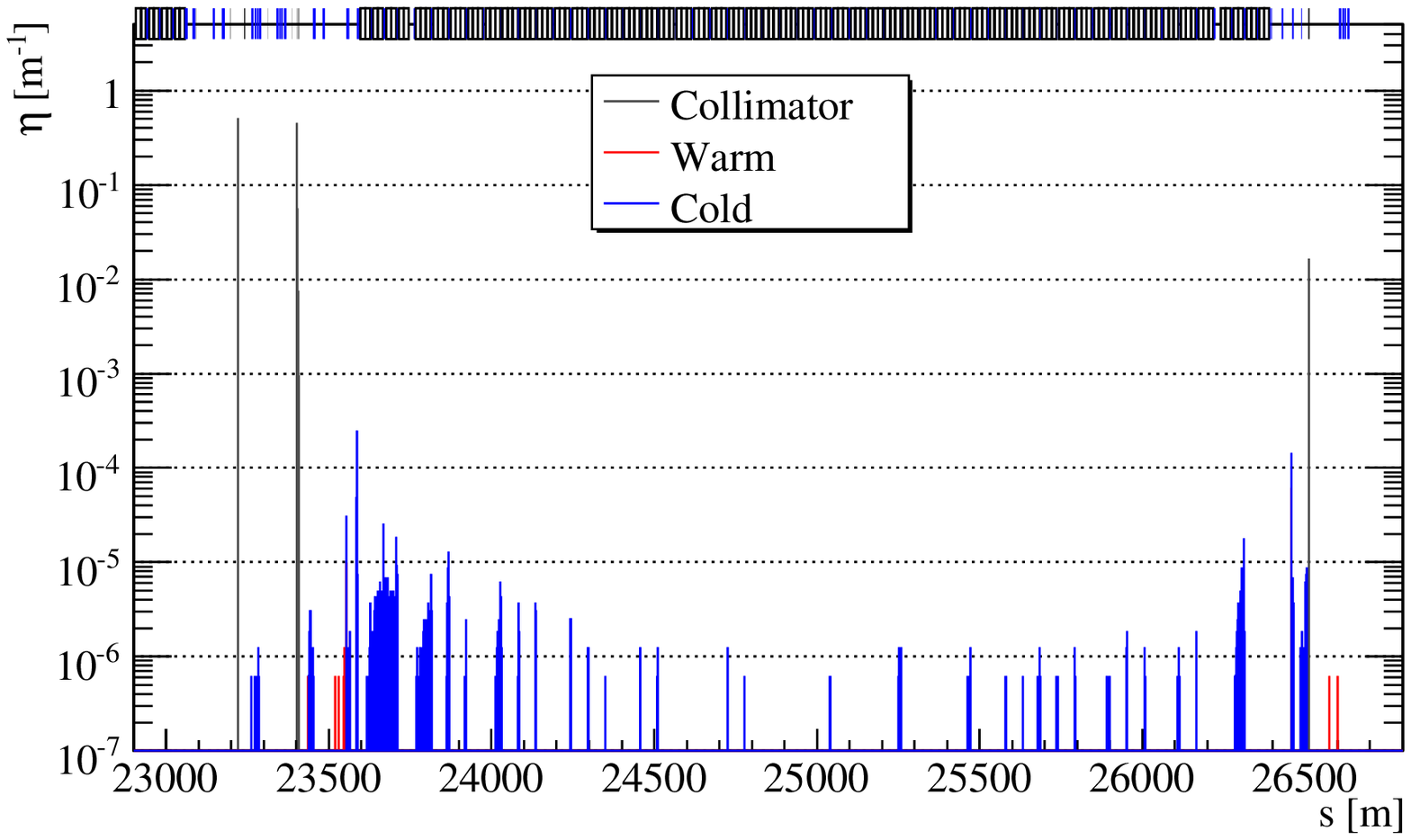}\label{fig:IR8_5s_zoom}}
\caption{Simulated beam-loss pattern for the IR8 layout with $\mathrm{Cry_1}$ at $5~\sigma$. The whole LHC (a), zoom of the arc 81 (b) (1~p~=~6.2$\times10^{-7}~\mathrm{m^{-1}}$).
}
\label{fig:IR8_5s}
\vspace*{-\baselineskip}
\end{figure}

\begin{figure}[t]
   \centering
   \includegraphics[width=90mm]{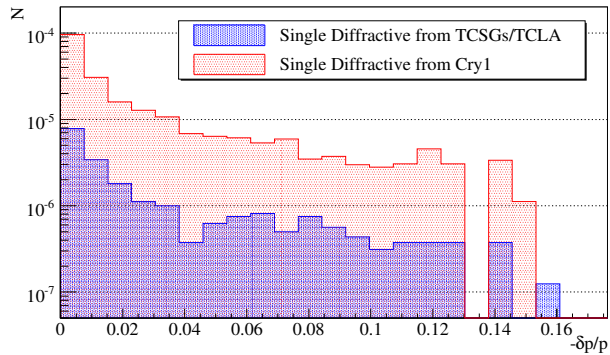}
   \caption{Spectrum of proton lost in the arc 81 due to a single diffractive interaction in the $\mathrm{Cry_1}$ (red) or in the TCSGs/TCLA in IR8 (blue), for the simulated beam loss pattern shown in Fig.~\ref{fig:IR8_5s}.}
   \label{fig:IR8_dpop}
   \vspace*{-\baselineskip}
\end{figure}


\begin{figure}[!t]
\subfloat[]{\includegraphics*[width=90mm]{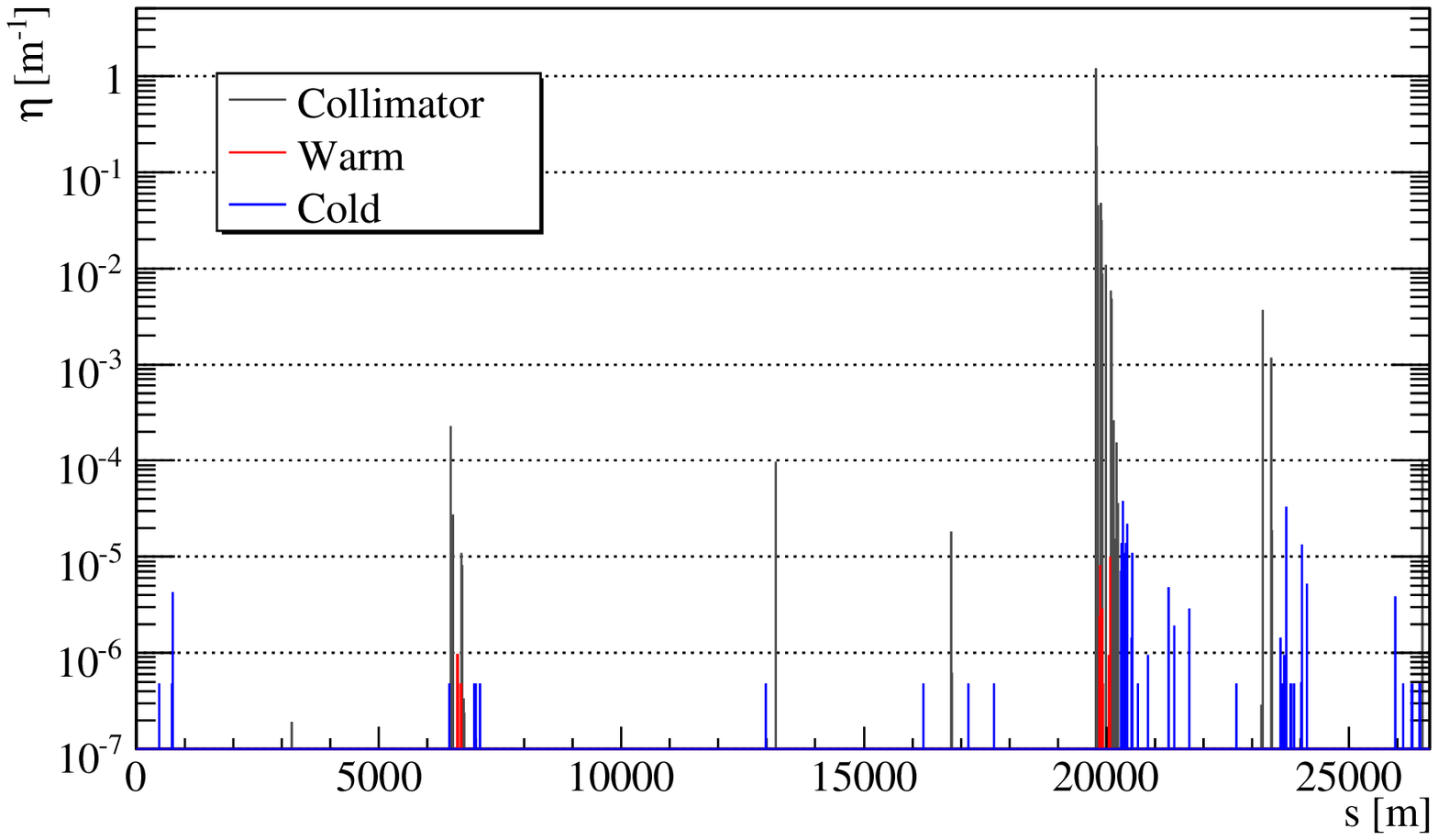}\label{fig:IR8_6s_tot}}\vspace*{-\baselineskip}\\
\vspace*{-\baselineskip}
\subfloat[]{\includegraphics*[width=90mm]{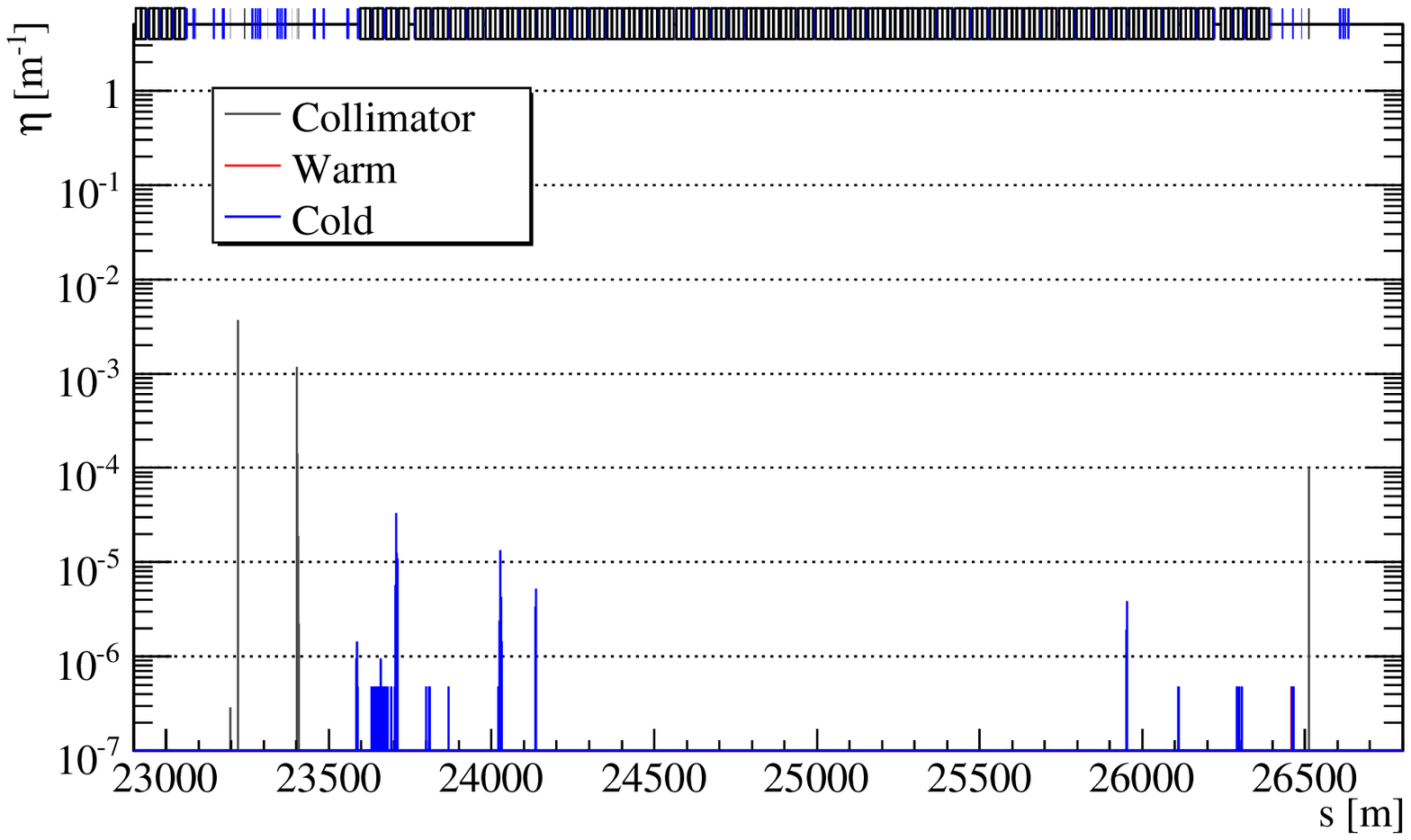}\label{fig:IR8_6s_zoom}}
\caption{Simulated beam-loss pattern for the IR8 layout with $\mathrm{Cry_1}$ at $6~\sigma$. The whole LHC (a), zoom of the arc 81 (b) (1~p~=~4.8$\times10^{-7}~\mathrm{m^{-1}}$).
}
\label{fig:IR8_6s}
\vspace*{-\baselineskip}
\end{figure}

\subsection{IR8 layout performance}
\label{sec:IR8_perf}

The beam loss pattern obtained placing $\mathrm{Cry_1}$ at $5~\sigma$ (i.e. same aperture of the TCP) is shown in Fig.~\ref{fig:IR8_5s}. The aperture of TCSGs and TCLA in IR8 is $10~\sigma$ and $13~\sigma$, respectively, in order to not interfere with the multi-turn betatron cleaning process. This layout and setting lead to an unacceptable loss pattern, as clearly visible from Fig.~\ref{fig:IR8_5s_zoom}, with very high cold losses in the arc 81 (much above the IR7-DS).

Studies to understand the source of these losses were performed. Most of them are due to off-momentum protons generated by the interaction with the $\mathrm{Cry_1}$ (i.e. single diffractive events), as shown in Fig.~\ref{fig:IR8_dpop}. These protons emerge from the $\mathrm{Cry_1}$ with a deflection that is not enough to be intercepted by the absorbers in IR8 and are lost at the dispersive peaks because of the momentum offset acquired.

Very high losses are also induced on the TCTPs placed in front of IP1 ($\sim\times200$ larger than operational), with a possible impact on ATLAS background.

Thus, it is not possible to use this layout and settings for $parasitic$ operations with full machine. The maximum peak in Fig.~\ref{fig:IR8_5s_zoom} is about 10 times larger than in Fig.~\ref{fig:pres_coll_IR7}. Hence, the circulating intensity that would lead to comparable loads on cold elements with respect to standard physics operations, is about 10 times smaller. Meaning that this configuration could be used only in $dedicated$ operations with a maximum of about 250 bunches of $1.1\times10^{11}$ protons circulating in the machine.

Simulations with settings of TCSGs-TCLA in IR8 down to 6-9~$\sigma$ were carried out and no significant changes in the loss pattern was observed. This is because the angular cut performed by the TCSGs in IR8 is not enough to intercept a significant fraction of single diffractive events coming from the $\mathrm{Cry_1}$. This limitation cannot be easily mitigated because of the proximity to IP8, which leads to large $\beta_y$ at the absorbers (i.e. large gap in mm that defines the angular cut). Simulations have been performed moving the absorbers in front of the main quadrupole in cell 6 right of IP8, which provides a much better angular cut thanks to the reduced $\beta_y$ value. However, losses on cold elements and TCTPs upstream of IP1 remain a substantial limitation.

\begin{figure}[t]
   \centering
   \includegraphics[width=90mm]{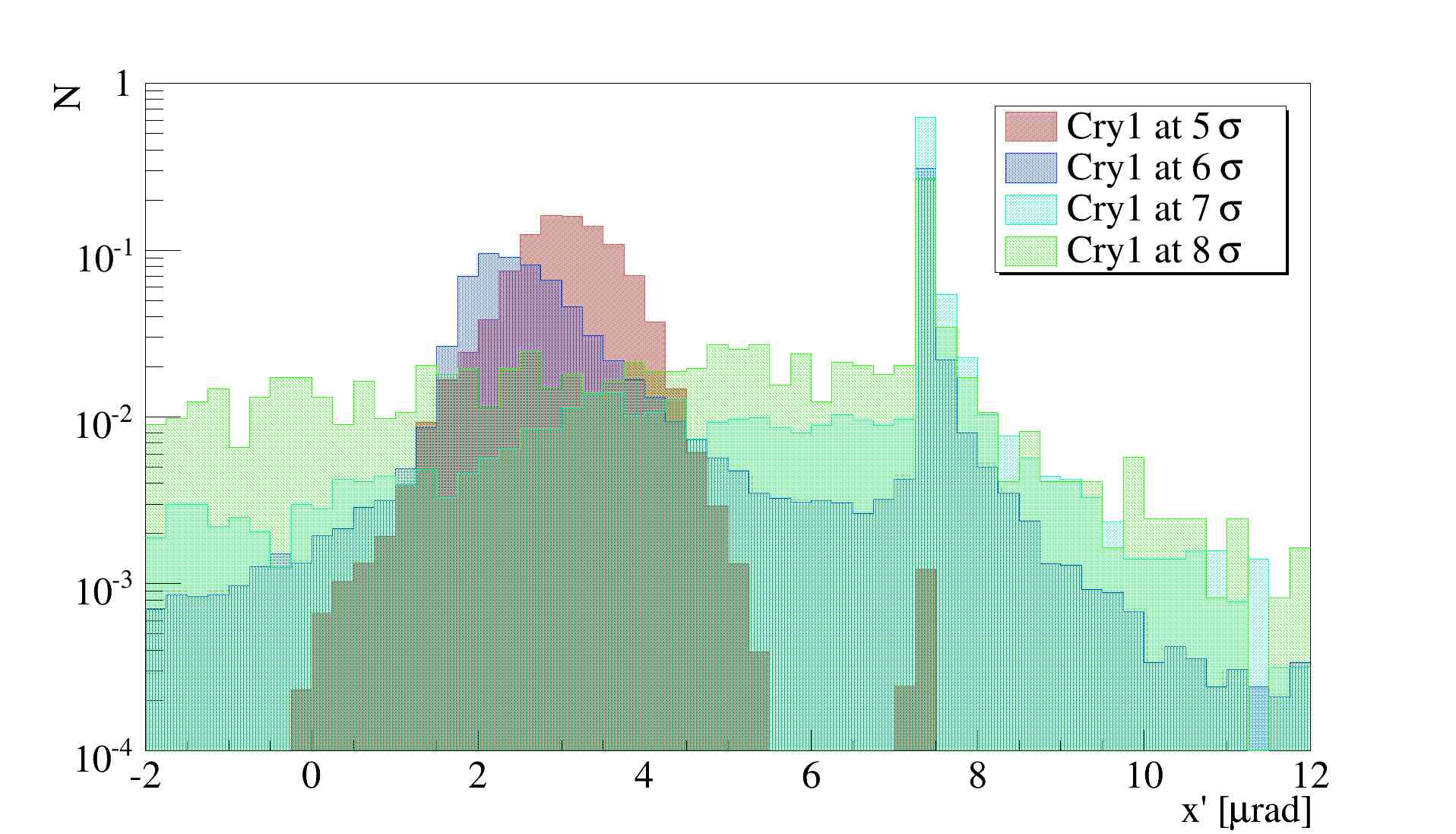}
   \caption{Angular distribution (in the machine reference frame) of impacting protons for different settings of $\mathrm{Cry_1}$ in IR8.}
   \label{fig:IR8_dist}
   \vspace*{-\baselineskip}
\end{figure}

\begin{figure}[t]
   \centering
   \includegraphics[width=90mm]{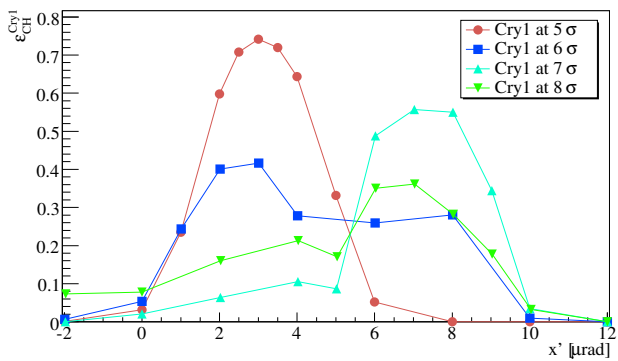}
   \caption{Channeling efficiency as a function of the crystal orientation (in the machine reference frame) for different settings of $\mathrm{Cry_1}$ in IR8.}
   \label{fig:IR8_eff}
   \vspace*{-\baselineskip}
\end{figure}

On the other hand, a significant reduction of losses is obtained for larger aperture of $\mathrm{Cry_1}$. Simulations with $\mathrm{Cry_1}$ settings from $6~\sigma$ to $8~\sigma$ in steps of $1~\sigma$ were carried out. A similar loss pattern as standard operational performance, in terms of cold losses, is obtained when $\mathrm{Cry_1}$ is at $6~\sigma$, as can be seen in Fig.~\ref{fig:IR8_6s}. Thus, $parasitic$ operations with $\mathrm{Cry_1}$ at $6~\sigma$ may be possible from collimation aspects. Of course, the larger the $\mathrm{Cry_1}$ aperture, the larger the angular distribution of impacting protons, as shown in Fig.~\ref{fig:IR8_dist}. Thus, angular scans of the $\mathrm{Cry_1}$ were simulated in order to find the optimal orientation at each setting, which are reported in Fig.~\ref{fig:IR8_eff}. The orientation leading to the maximum channeling efficiency (defined below) was adopted to produce all the results reported here. The peak in Fig.~\ref{fig:IR8_dist} at 7-8~$\mu$rad is due to emerging protons from the TCPs that hit $\mathrm{Cry_1}$ in the same turn.

Let us define the number of proton on target (PoT) as figure of merit to evaluate the system performance. The number of PoT can be estimated as:

\begin{equation}
PoT(t)=\frac{1}{2}\frac{I(t)}{\tau}\exp(-\frac{t}{\tau})~\frac{N_{\rm{imp}}^{\mathrm{Cry_1}}}{N_{\rm{sim}}}~\varepsilon_{CH}^{\mathrm{Cry_1}}\:,
\end{equation}

\begin{table*}[t]
\setlength\tabcolsep{10pt}
   \centering
   \caption{Fraction of simulated protons that hit $\mathrm{Cry_1}$ in both layouts and relative channeling efficiency, together with integrated PoT in a 10 h fill with 200 h beam lifetime.}
   \begin{tabular}{c|ccc|ccc}
       \hline\noalign{\smallskip}
       & & \textbf{IR3} & & &  \textbf{IR8}  \\
       \textbf{$\mathrm{Cry_1}$} settings & \textbf{$\frac{N_{\rm{imp}}^{\mathrm{Cry_1}}}{N_{\rm{sim}}}$} & $\varepsilon_{CH}^{\mathrm{Cry_1}}$ & $\int_{10h} PoT(t) dt$ [p]& \textbf{$\frac{N_{\rm{imp}}^{\mathrm{Cry_1}}}{N_{\rm{sim}}}$} & $\varepsilon_{CH}^{\mathrm{Cry_1}}$ & $\int_{10h} PoT(t) dt$ [p]\\
       \noalign{\smallskip}\hline\noalign{\smallskip}
           5         & 0.78            & 0.66     &  $2.8\times10^{12}$ & 0.67            & 0.74    &  $2.7\times10^{11}$ \\ 
           6         & $2.4\times10^{-3}$            & 0.40 &  $5.2\times10^{9}$ & $3.1\times10^{-3}$            & 0.41 &  $6.9\times10^{9}$\\ 
           7	& $2.7\times10^{-4}$		& 0.26 &  $3.8\times10^{8}$ & $3.5\times10^{-4}$		& 0.56 &  $1.1\times10^{9}$\\
           8	& $1.3\times10^{-4}$		& 0.12	&  $8.4\times10^{7}$	 & $5.3\times10^{-5}$		& 0.36	&  $1.0\times10^{8}$\\	
       \noalign{\smallskip}\hline
   \end{tabular}
   \label{tab:IR3_8_fact}
   \vspace*{-\baselineskip}
\end{table*}



\begin{figure}[!t]
\subfloat[]{\includegraphics*[width=90mm]{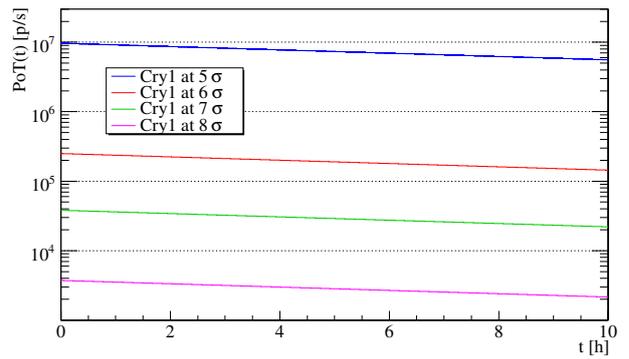}\label{fig:PoT_IR8_inst}}\vspace*{-\baselineskip}\\
\vspace*{-\baselineskip}
\subfloat[]{\includegraphics*[width=90mm]{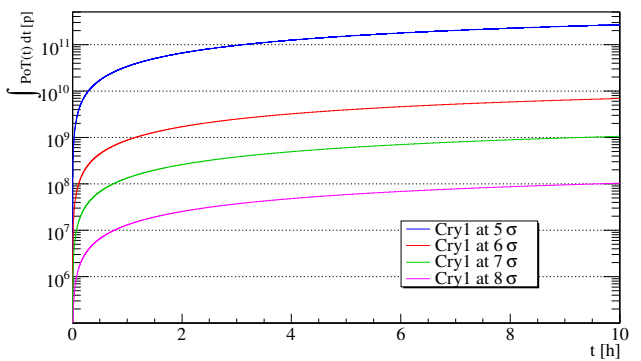}\label{fig:PoT_IR8_int}}
\caption{Expected instantaneous (a) and integrated (b) PoT for different settings of $\mathrm{Cry_1}$ in IR8 during a 10~h fill with 200~h beam lifetime.}
\label{fig:PoT_IR8}
\vspace*{-\baselineskip}
\end{figure}

\noindent where $\frac{I(t)}{\tau}\exp(-\frac{t}{\tau})$ is the total beam loss rate for a certain beam lifetime $\tau$ and circulating intensity $I(t)$, $\frac{1}{2}$ is the sharing of the total loss rate between the horizontal and vertical planes~\cite{BelenLosses}, $\frac{N_{\rm{imp}}^{\mathrm{Cry_1}}}{N_{\rm{sim}}}$ is the fraction of simulated protons that hit the $\mathrm{Cry_1}$, and $\varepsilon_{CH}^{\mathrm{Cry_1}}=\frac{N_{CH}^{\mathrm{Cry_1}}}{N_{\rm{imp}}^{\mathrm{Cry_1}}}$ is the channeling efficiency of $\mathrm{Cry_1}$ (i.e. fraction of impacting protons trapped between crystalline planes for the entire path in the crystal). The reduction of circulating protons due to the collisions in the 4 IPs can be approximated as:

\begin{equation}
I(t)=I_{\rm{tot}}\exp(-\frac{t}{\tau_{\rm{BO}}})\:,
\end{equation}

\noindent where $I_{\rm{tot}}$ is the total stored intensity at the beginning of the fill, while $\exp(-\frac{t}{\tau_{\rm{BO}}})$ takes into account the intensity decay due to burn-off (with $\tau_{\rm{BO}}\sim20$~h~\cite{BOEvian2019}). A summary of what has been obtained for different $\mathrm{Cry_1}$ settings is reported in Table~\ref{tab:IR3_8_fact}. Assuming a beam lifetime of $\tau\sim200$~h according to usual operational values in 2018~\cite{BelenLifetime}, the achievable instantaneous and integrated $PoT(t)$ during one fill are shown in Fig.~\ref{fig:PoT_IR8}. The integrated PoT in 10 h (usual fill length during LHC operations) are reported in Table~\ref{tab:IR3_8_fact}. The maximum $I_{\rm{tot}}$ stored in 2018 was of 2556 bunches with about $1.1\times10^{11}$ protons per bunch, which is equivalent to about $2.8\times10^{14}$ protons injected~\cite{BelenLifetime}. This initial intensity is scaled by a factor 10 for $\mathrm{Cry_1}$ setting of $5~\sigma$, as explained previously to allow \emph{dedicated} operations.

\begin{figure}[!t]
\subfloat[]{\includegraphics*[width=90mm]{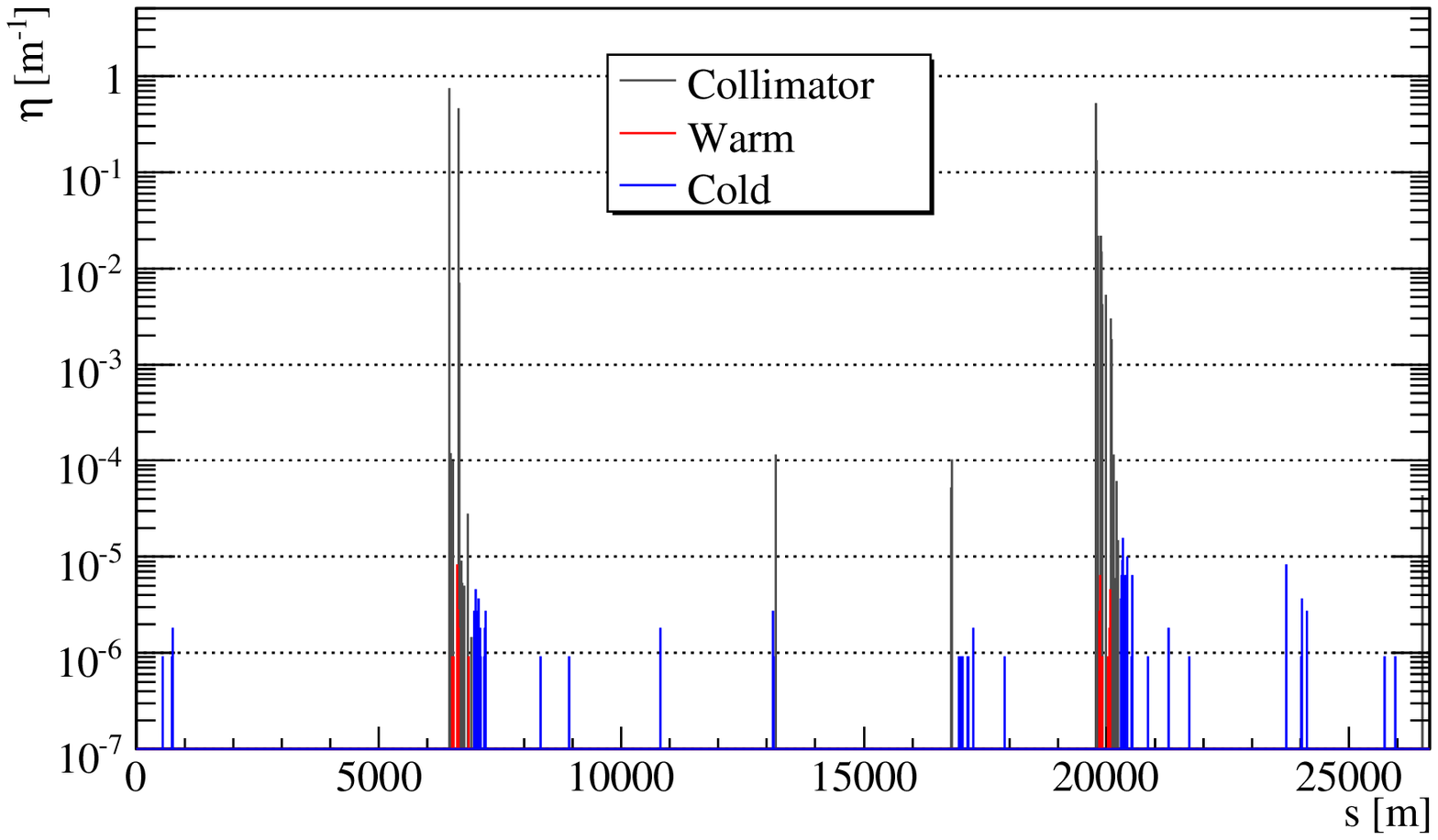}\label{fig:IR3_5s_tot}}\vspace*{-\baselineskip}\\
\vspace*{-\baselineskip}
\subfloat[]{\includegraphics*[width=90mm]{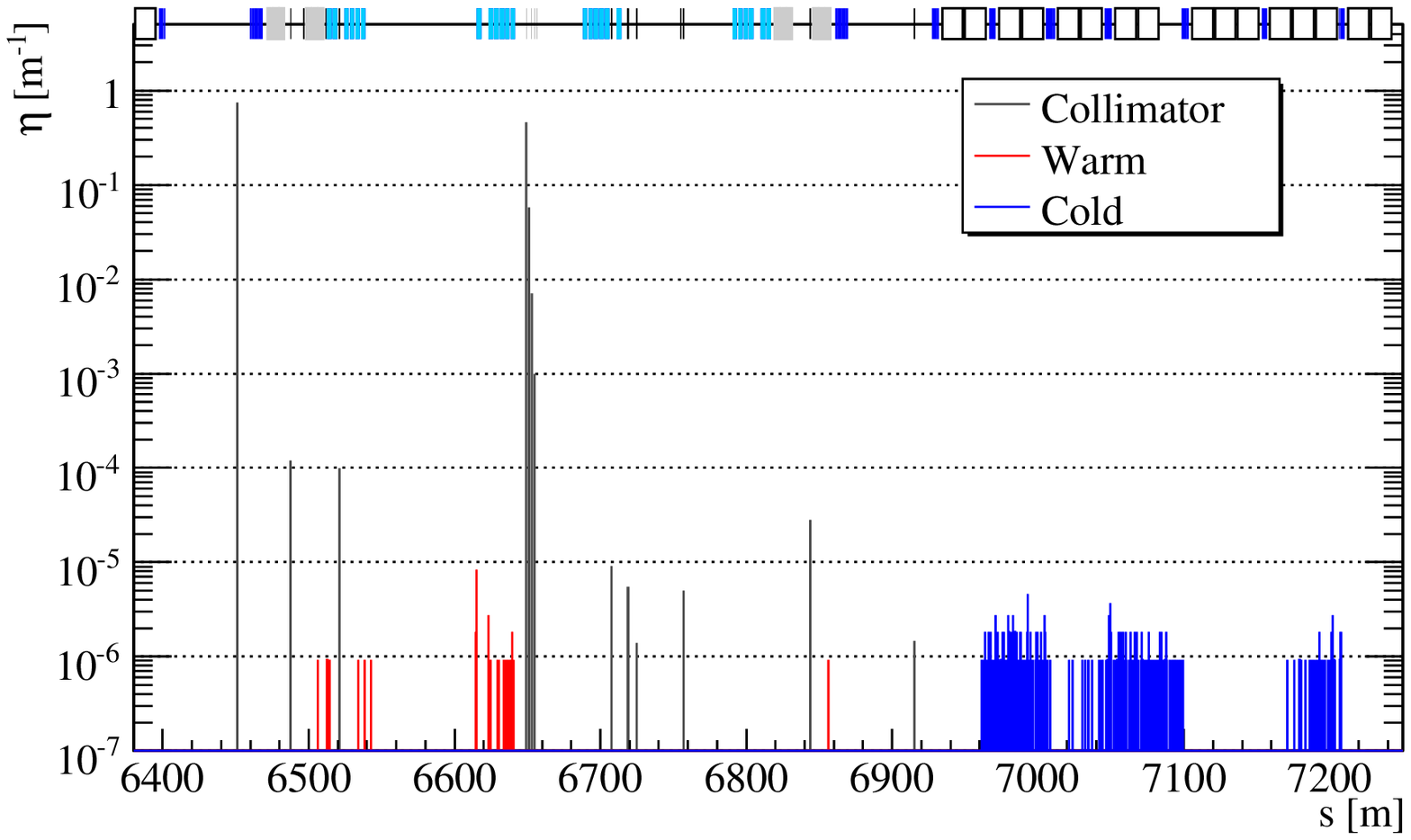}\label{fig:IR3_5s_zoom}}
\caption{Simulated beam-loss pattern for the IR3 layout with $\mathrm{Cry_1}$ at $5~\sigma$. The whole LHC (a), zoom of the IR3 insertion (b) (1~p~=~9.1$\times10^{-7}~\mathrm{m^{-1}}$).
}
\label{fig:IR3_5s}
\vspace*{-\baselineskip}
\end{figure}

\subsection{IR3 layout performance}

\begin{figure}[!t]
   \centering
   \includegraphics[width=90mm]{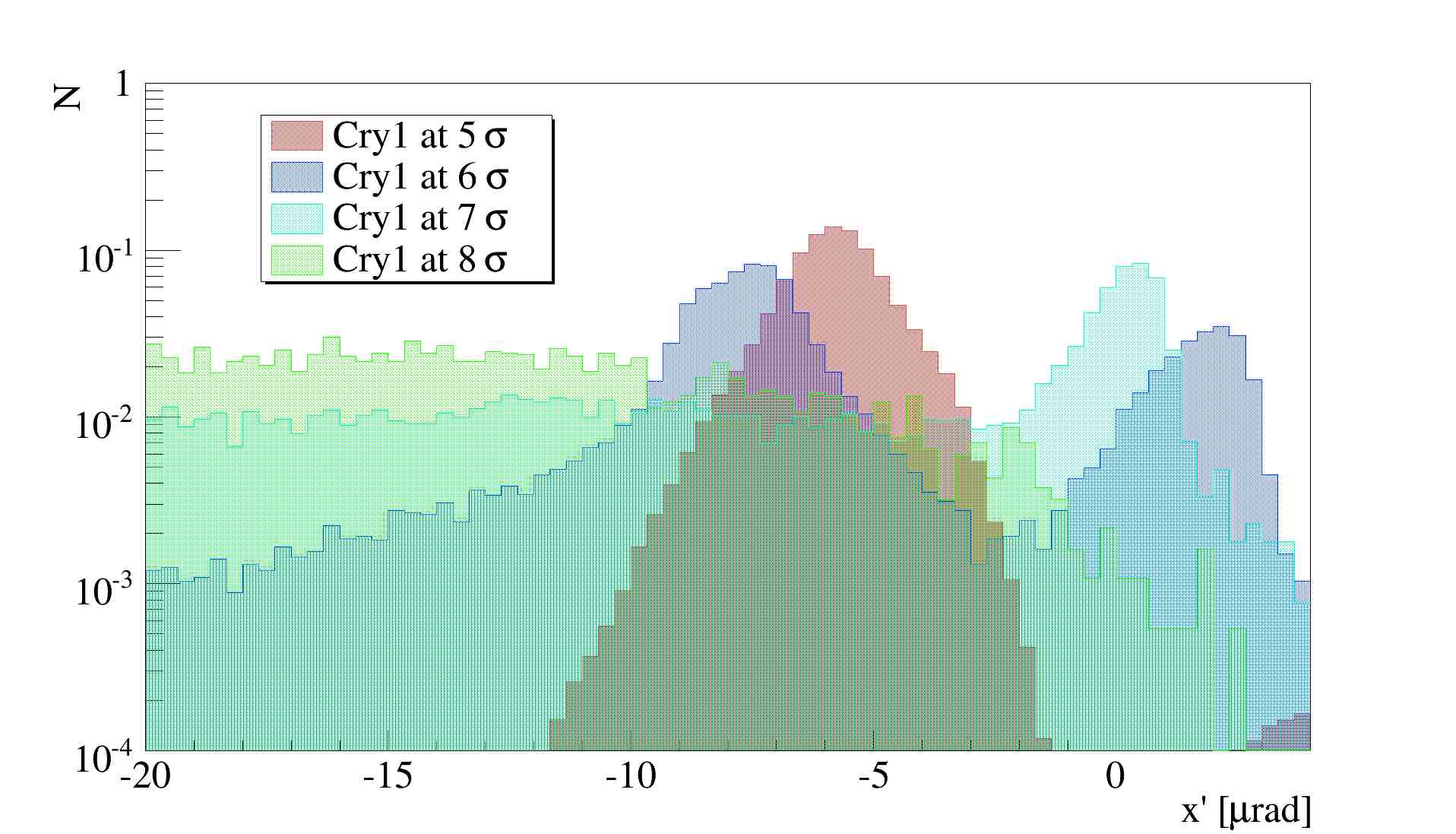}
   \caption{Angular distribution (in the machine reference frame) of impacting protons for different settings of the $\mathrm{Cry_1}$ in IR3.}
   \label{fig:IR3_dist}
   \vspace*{-\baselineskip}
\end{figure}

\begin{figure}[!t]
   \centering
   \includegraphics[width=90mm]{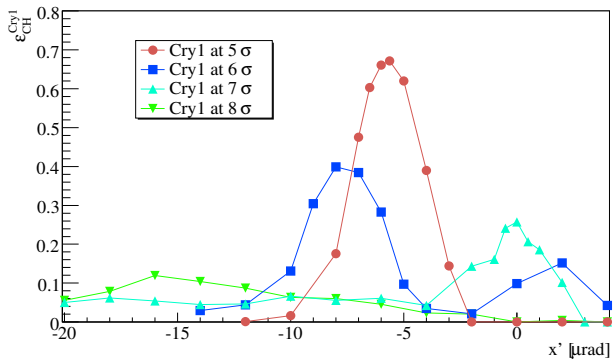}
   \caption{Channeling efficiency as a function of the crystal orientation (in the machine reference frame) for different settings of the $\mathrm{Cry_1}$ in IR3.}
   \label{fig:IR3_eff}
   \vspace*{-\baselineskip}
\end{figure}

\begin{figure}[t]
\subfloat[]{\includegraphics*[width=90mm]{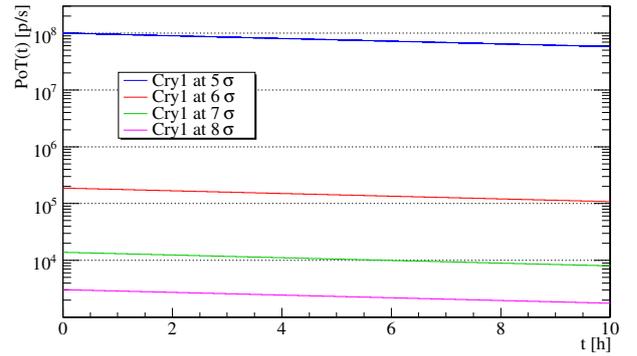}\label{fig:PoT_IR3_inst}}\vspace*{-\baselineskip}\\
\vspace*{-\baselineskip}
\subfloat[]{\includegraphics*[width=90mm]{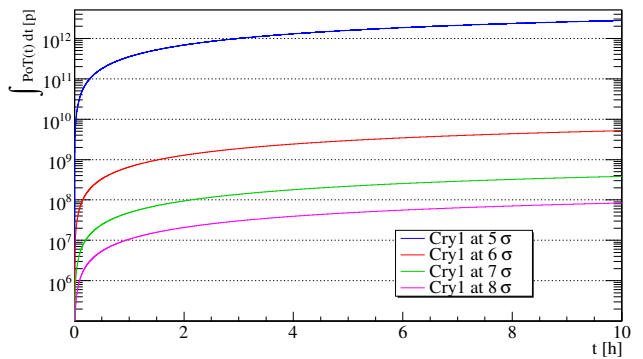}\label{fig:PoT_IR3_int}}
\caption{Expected instantaneous (a) and integrated (b) PoT for different settings of $\mathrm{Cry_1}$ in IR3 during a 10~h fill with 200~h beam lifetime.
}
\label{fig:PoT_IR3}
\vspace*{-\baselineskip}
\end{figure}

The beam loss pattern obtained with $\mathrm{Cry_1}$ at $5~\sigma$ (i.e. same aperture of the TCP) is shown in Fig.~\ref{fig:IR3_5s}. The aperture of TCSGs and TCLA in IR3 is $10~\sigma$ and $13~\sigma$, respectively, in order to not interfere with the multi-turn betatron cleaning process. In principle this layout would allow $parasitic$ operations during standard LHC operations also with such a tight $\mathrm{Cry_1}$ setting. The loss pattern in Fig.~\ref{fig:IR3_5s} that do not show any peak of cold losses above the threshold defined by Fig.~\ref{fig:pres_coll_IR7}. This is mainly due to the reduced bending and length needed for the $\mathrm{Cry_1}$ and the tighter angular cut performed by the TCSGs in IR3, with respect to the IR8 layout. Thus, less protons experience nuclear interactions in the $\mathrm{Cry_1}$ itself and are intercepted by the TCSGs more efficiently. Nevertheless, simulations with different $\mathrm{Cry_1}$ aperture were carried out using the same procedure described in section~\ref{sec:IR8_perf}. The angular distribution of impacting protons for $\mathrm{Cry_1}$ at different settings and the expected channeling efficiency are shown in Fig.~\ref{fig:IR3_dist} and~\ref{fig:IR3_eff}, respectively. The achievable instantaneous and integrated $PoT(t)$ during one fill are shown in Fig.~\ref{fig:PoT_IR3}. A summary is reported in Table~\ref{tab:IR3_8_fact}. 



\vspace{-4mm}

\section{$\mathrm{\Lambda_c}$ yield performance}
\label{sec:sec_cry}

\vspace{-2mm}

Several considerations are required to optimize the yield of $\mathrm{\Lambda_c}$ channeled by $\mathrm{Cry_2}$ and the measurement of their precession. The spin precession angle ($\phi$) of $\mathrm{\Lambda_c}$ in bent crystals is proportional to their bending ~\cite{lyuboshits1979spin}:

\begin{equation}
\phi=\left( 1 + \gamma\, \frac{g-2}{2} \right) \, \theta_b^{\mathrm{Cry_2}}\:,
\end{equation}

%

\noindent where $\gamma=\frac{E}{m}$ with $E$ and $m$ energy and rest mass of the particle, $\theta_b^{\mathrm{Cry_2}}$ is the bending angle of $\mathrm{Cry_2}$ and $g$ the gyromagnetic factor (or dimensionless magnetic moment).
Thus, the larger the crystal bending and the $\mathrm{\Lambda_c}$ energy, the larger the induced precession (i.e. the easier its measurement). Bending of the order of mrad is needed to obtain reasonable precision on $\phi$, and hence on the gyromagnetic factor $g$ at LHC energy~\cite{bagli2017electromagnetic,fomin2017feasibility}.

Such large bending angles require long crystals with large bending radius $(R)$. Large bending radius are needed because of the critical bending radius $(R_c)$, which is the radius where the potential well between crystalline planes disappears and the channeling regime is no longer possible. In particular, $R_c$ depends linearly on the particle energy $(pv)$ as~\cite{Bir,tsyganov1976some}:

\begin{equation} \label{eq:crit_rad}
R_{c}=\frac{pv}{U'_{max}}\:, 
\end{equation}
\noindent where $U'_{max}$ is the maximum gradient of interplanar electric potential. For the plane (110) in silicon crystal $U'_{max}\approx6$~GeV/cm, thus for 6.5~TeV positively charged particles $R_c\approx11$~m. This implies an angular acceptance of the channeling process in bent crystal $(\theta_{c}^{b})$ of:

%
%

\begin{equation} \label{eq:crit_ang_bnd}
\theta_{c}^{b}=\theta_{c}\left( 1 -\frac{R_{c}}{R}\right)=\sqrt{\frac{2U_{max}}{pv}}\left( 1 -\frac{R_{c}}{R}\right)\:,
\end{equation}

\noindent where $\theta_{c}$ is the critical channeling angle in straight crystal~\cite{Lind} and $U_{max}\approx$~21.3~eV in Si crystals~\cite{gemmell1974channeling} is the maximum of the potential well between crystalline planes.


On the other hand, the longer the crystal, the higher the probability of dechanneling and nuclear interactions. 
The dechanneling process can be described as an exponential decay of the initial population of channeled particles. Using diffusion theory it is possible to derive the contribution given by interactions with electrons in the crystalline channel \cite{Bir}, leading to the \emph{characteristic electronic dechanneling length}, which is linear in particle energy and can be written as:

\begin{equation} \label{eq:dech_len}
L_{D}^{e}=\frac{256}{9\pi^{2}}\frac{pv}{\ln(2m_{e}c^{2}\gamma/I)-1}\frac{a_{TF}d_{p}}{Z_{i}r_{e}m_{e}c^{2}} \:,
\end{equation}

\noindent where $I$ is the ionisation potential ($I\simeq172$ eV in Si), $Z_{i}$ the electric charge of the channeled particle with its relativistic factor $\gamma$, while $r_{e}$ and $m_{e}$ are the classical radius and rest mass of the electron, respectively, and $a_{TF}$ and $d_{p}$ are the Thomas-Fermi constant and interplanar distance, respectively. However,  electronic dechanneling only describes a ``slow" dechanneling regime (i.e. order of cm), due to the very small variation in momentum from scattering with electrons in the channel, leading to an incomplete treatment of the whole process. Hard scattering on nuclei can lead to ``fast" dechanneling as a result of single interactions  (i.e. order of mm). Therefore, a \emph{characteristic nuclear dechanneling length} $(L_{D}^{N})$ must also be taken into account for a reliable parameterization of the entire dechanneling process. Analytical treatments of this process are not available in literature, but this characteristic length for nuclear dechanneling can be derived by  appropriate scaling of the electronic value~\cite{mirarchi2015crystal_rout,routineYellow}. Thus, when the crystal length becomes comparable to the dechanneling length, a significant fraction on channeled particles will escape from the crystalline planes without acquiring the deflection (i.e. precession) required and will not be in the detector acceptance.

The combined effect of dechanneling and angular acceptance makes a bent crystal behaving as a spectrometer over the energy spectrum of impacting particles. Qualitatively, one can expect a linear increase of efficiency as a function of energy because of the growing $L_{D}=L_{D}^{e}+L_{D}^{N}$, which is followed by a plateau and a new decrease of efficiency due to the reduction of $\theta_{c}^{b}$. Thus, the smaller the bending angle, the larger the maximum channeling efficiency, while the longer the crystal, the larger the energy range with stable channeling efficiency.

\begin{figure}[!t]
   \centering
   \includegraphics[width=85mm]{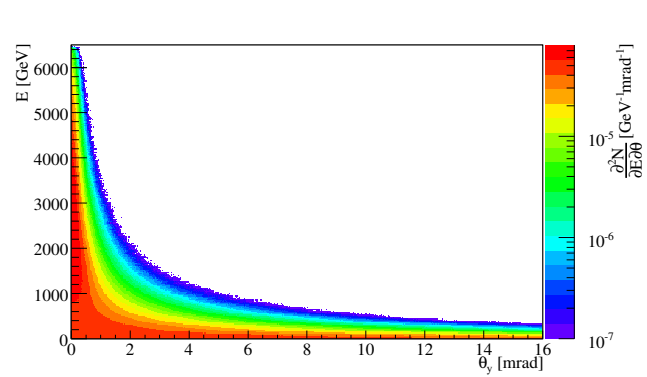}
   \caption{Distribution of $\mathrm{\Lambda_c}$ for 6.5 TeV protons on target, obtained using \textsc{Pythia8.240}.}
   \label{fig:Lc_spec}
   \vspace*{-\baselineskip}
\end{figure}

Let us define the channeling efficiency of $\mathrm{Cry_2}$, without considering the $\mathrm{\Lambda_c}$ decay, as:

\begin{equation}
\varepsilon_{CH}^{\mathrm{Cry_2}}(E)=\frac{N_{CH}^{\mathrm{\Lambda_c}}(E)}{N_{\rm{imp}}^{\mathrm{\Lambda_c}}(E)}\:,
\label{eq:cry2_eff}
\end{equation}

\noindent where $N_{CH}^{\mathrm{\Lambda_c}}(E)$ is the number of $\mathrm{\Lambda_c}$ that remain in channeling for the full crystal length, and $N_{\rm{imp}}^{\mathrm{\Lambda_c}}(E)$ is the number of impacting $\mathrm{\Lambda_c}$.

The distribution of $N_{\rm{imp}}^{\mathrm{\Lambda_c}}(E)$ was obtained using the \textsc{Pythia8.240} event generator, starting from the impacting distribution of protons on the target, coming from $\mathrm{Cry_1}$. The distribution of $\mathrm{\Lambda_c}$ obtained for 6.5 TeV protons on target is shown in Fig.~\ref{fig:Lc_spec}.

\begin{figure}[t]
   \centering
   \includegraphics[width=90mm]{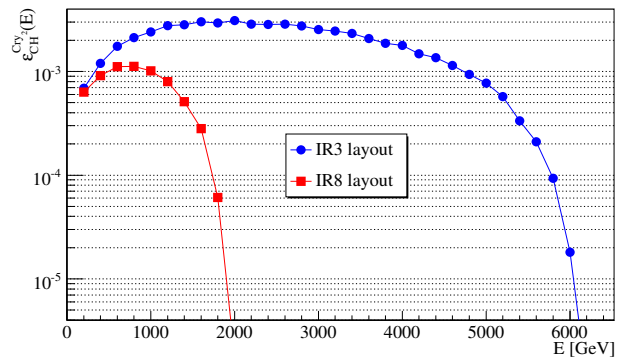}
   \caption{Channeling efficiency of $\mathrm{Cry_2}$ as defined in equation~(\ref{eq:cry2_eff}).}
   \label{fig:cry2_eff}
\end{figure}

\begin{figure}[t]
   \centering
   \includegraphics[width=90mm]{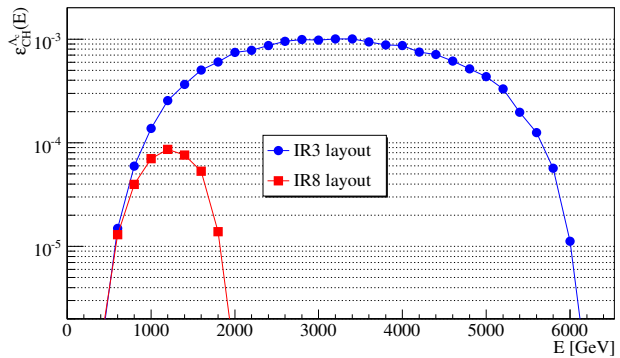}
   \caption{Efficiency of $\mathrm{\Lambda_c}$ channeled by $\mathrm{Cry_2}$ as defined in equation~(\ref{eq:Lc_eff}).}
   \label{fig:Lc_eff}
\end{figure}

Transverse dimensions of $\mathrm{Cry_2}$, compared to the impacting distribution of protons on target, are crucial in order to intercept the maximum number of produced $\mathrm{\Lambda_c}$ (i.e. $\mathrm{Cry_2}$ thickness must be larger than the impacting distribution on target) while ensuring a uniform crystal curvature. In particular, a ratio of $\frac{R}{y}\sim3000$ was used, where $R$ and $y$ are bending radius and crystal thickness, respectively. Thus, $y_{Cry_{2}}^{IR8}=2$~mm and $y_{Cry_{2}}^{IR3}=5$~mm. The $\varepsilon_{CH}^{\mathrm{Cry_2}}(E)$ for the two layouts are shown in Fig.~\ref{fig:cry2_eff}.

This efficiency must be convoluted with the $\mathrm{\Lambda_c}$ decay inside the crystal. Thus, let us define  efficiency of $\mathrm{\Lambda_c}$ channeled by the $\mathrm{Cry_2}$ as: 

\begin{equation}
\varepsilon_{CH}^{\mathrm{\Lambda_c}}(E)=\varepsilon_{CH}^{\mathrm{Cry_2}}(E)\exp(-l_{\mathrm{Cry_2}}/c\gamma\tau_{\mathrm{\Lambda_c}})\:,
\label{eq:Lc_eff}
\end{equation}

\noindent where $l_{\mathrm{Cry_2}}$ is the length of $\mathrm{Cry_2}$ depending on the layout, while $c\tau_{\mathrm{\Lambda_c}}=59.9 \mu$m~\cite{PDGLc} is the decay length of the ${\rm\Lambda_c}$ baryon. The efficiency of ${\rm\Lambda_c}$ channeled by $\mathrm{Cry_2}$ is shown in Fig.~\ref{fig:Lc_eff}.



A further step towards the evaluation of expected $\mathrm{\Lambda_c}$ yield is the convolution with the expected number of PoT and production in the target. The number of $\mathrm{\Lambda_c}$ produced can be calculated as:

\begin{equation}
N_{\mathrm{\Lambda_c}}=N_A\rho_{\rm{t}} l_{\rm{t}} \sigma(\mathrm{\Lambda_c}) = 0.6\times10^{-4} \mathrm{\Lambda_c}/p\:,
\end{equation}

\noindent where $N_A$, $\rho_{\rm{t}}$, $l_{\rm{t}}$ and $\sigma(\mathrm{\Lambda_c})$ are the Avogadro's number, target density, target length, and total cross section, respectively. In particular $\sigma(\mathrm{\Lambda_c})=10.13~\mu\mathrm{b}$ was used for 6.5 TeV impacting protons, as calculated using \textsc{Pythia8.240}. Note that the $\mathrm{\Lambda_c}$ production and decay in the target volume must be taken into account as:  

\begin{equation}
P_t(E)=\frac{1}{l_{\rm{t}}}\int_{0}^{l_{\rm{t}}} \exp(-l/c\gamma\tau_{\mathrm{\Lambda_c}}) dl\:.
\end{equation}

\begin{figure}[t]
   \centering
   \includegraphics[width=90mm]{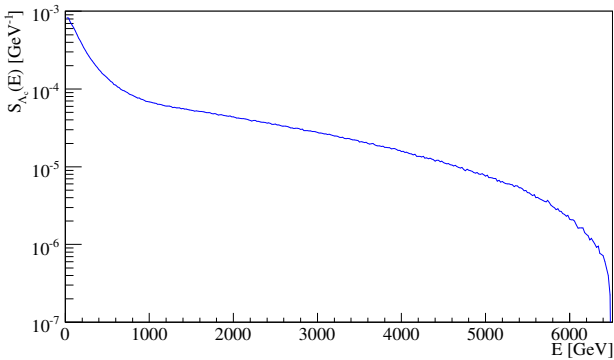}
   \caption{Spectrum of $\mathrm{\Lambda_c}$ produced by 6.5~TeV protons on target.}
   \label{fig:Lc_distr}
\end{figure}

 The production spectrum of $\mathrm{\Lambda_c}$ $(S_{\mathrm{\Lambda_c}}(E))$ must be also taken into account and it is shown in Fig.~\ref{fig:Lc_distr}.

In conclusion, the yield of $\mathrm{\Lambda_c}$ emerging from $\mathrm{Cry_2}$ that have acquired the desired precession can be expressed as:

 \begin{equation}
Y_{\mathrm{\Lambda_c}}(E)= N_{\mathrm{\Lambda_c}}P_t(E) S_{\mathrm{\Lambda_c}}(E) \varepsilon_{CH}^{\mathrm{\Lambda_c}}(E)\int_{10h} PoT(t)dt\:.
\label{eq:Lc_yield}
\end{equation}

\begin{figure}[t]
   \centering
   \includegraphics[width=90mm]{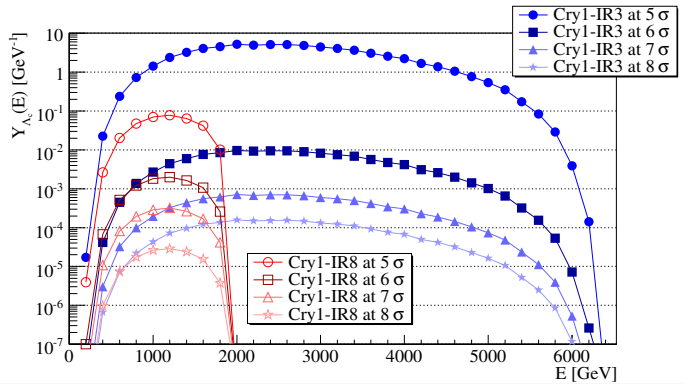}
   \caption{Expected yield of $\mathrm{\Lambda_c}$ emerging from $\mathrm{Cry_2}$ that acquired the desired precession in 10 h LHC fill as defined in Eq.~(\ref{eq:Lc_yield}).}
   \label{fig:Lc_yield}
\end{figure}

Finally, the $Y_{\mathrm{\Lambda_c}}(E)$ for the two layouts proposed and possible operational scenario are shown in Fig.~\ref{fig:Lc_yield}, using $\int_{10h} PoT(t)dt$ reported in Tables~\ref{tab:IR3_8_fact} and~\ref{tab:IR3_8_fact}. 

%
%
%
%

The parameters of $\mathrm{Cry_2}$ for the IR3 layout were chosen to ensure:

\begin{itemize}

\item Smallest bending that lead to measurable precession and clear separation of the expected physics signal with respect to the background~\cite{fomin2017feasibility}.

\item Largest length that increases as much as possible the energy range of stable channeling efficiency, but avoids channeling of 6.5 TeV protons coming from the $\mathrm{Cry_1}$.

\end{itemize}

The flexibility provided by the IR3 layout allows to optimize these parameters leading to a significant gain with respect to the IR8 layout in terms of $\mathrm{\Lambda_c}$ yield, for every operational configuration and $\mathrm{Cry_1}$ settings considered, as clearly visible from Fig.~\ref{fig:Lc_yield}.

The channeling efficiency of the $\mathrm{Cry_2}$ was also calculated using a parameterization based on the Monte-Carlo simulation of particle propagation through a crystalline lattice, taking into account incoherent scattering on  electrons and thermal vibrations of the atoms at lattice nodes~\cite{FominThesis}. The results are in agreement at the level of few  \%.



\subsection{Optimized operational scenario}

Operations with $\mathrm{Cry_1}$ at $5~\sigma$ are likely to be excluded for both layouts, due to machine protection aspects. On the other hand, the larger the $\mathrm{Cry_1}$ setting, the smaller the expected rate of PoT. Thus, the best compromise between performance and machine protection constraints has to be found. Minimum setting of the $\mathrm{Cry_1}$ has to ensure that:

\begin{enumerate}

\item It is impossible that $\mathrm{Cry_1}$ becomes the primary collimation stage.

\item Local losses are always below safe limits even if the channeling orientation of $\mathrm{Cry_1}$ is lost.

\item Local losses are always below safe limits also in the occurrence of beam lifetime drops.

\end{enumerate}

Margins to ensure item \textbf{1} are defined by optics errors and orbit stability. Optics corrections in the LHC ensure a peak $\delta\beta/\beta<10\%$~\cite{tomas2012record,tomas2017review}, which corresponds to a $5\%$ error on beam size. Thus, $\mathrm{Cry_1}$ cannot be set below $5.5~\sigma$ if TCPs are set at $5~\sigma$. However, this margin can be reduced by performing a beam-based alignment of the $\mathrm{Cry_1}$ with respect to the TCPs, because the eventual $\delta\beta/\beta$ will no longer change after optics correction are deployed. Nevertheless, fill-to-fill orbit stability and reproducibility are also in the range of $<100~\mu$m~\cite{wenniger2016machine,belen2019machine}, which is also of the order of $0.5~\sigma$ at $\mathrm{Cry_1}$ locations. Thus, a lower limit of $5.5~\sigma$ for $\mathrm{Cry_1}$ setting is defined for the machine configuration taken into account.

\begin{figure}[!t]
\subfloat[]{\includegraphics*[width=90mm]{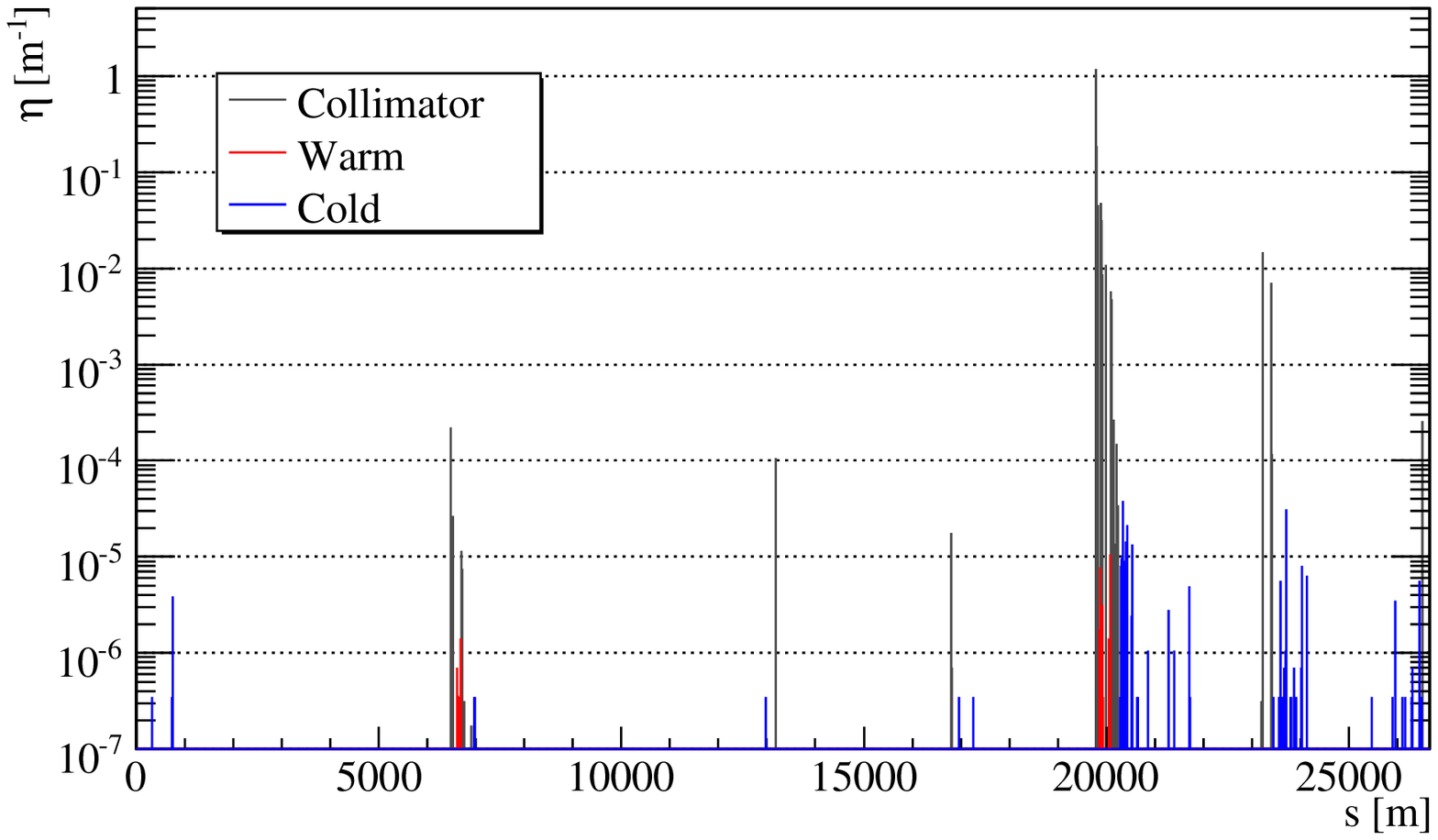}\label{fig:IR8_5p5s_tot}}\vspace*{-\baselineskip}\\
\vspace*{-\baselineskip}
\subfloat[]{\includegraphics*[width=90mm]{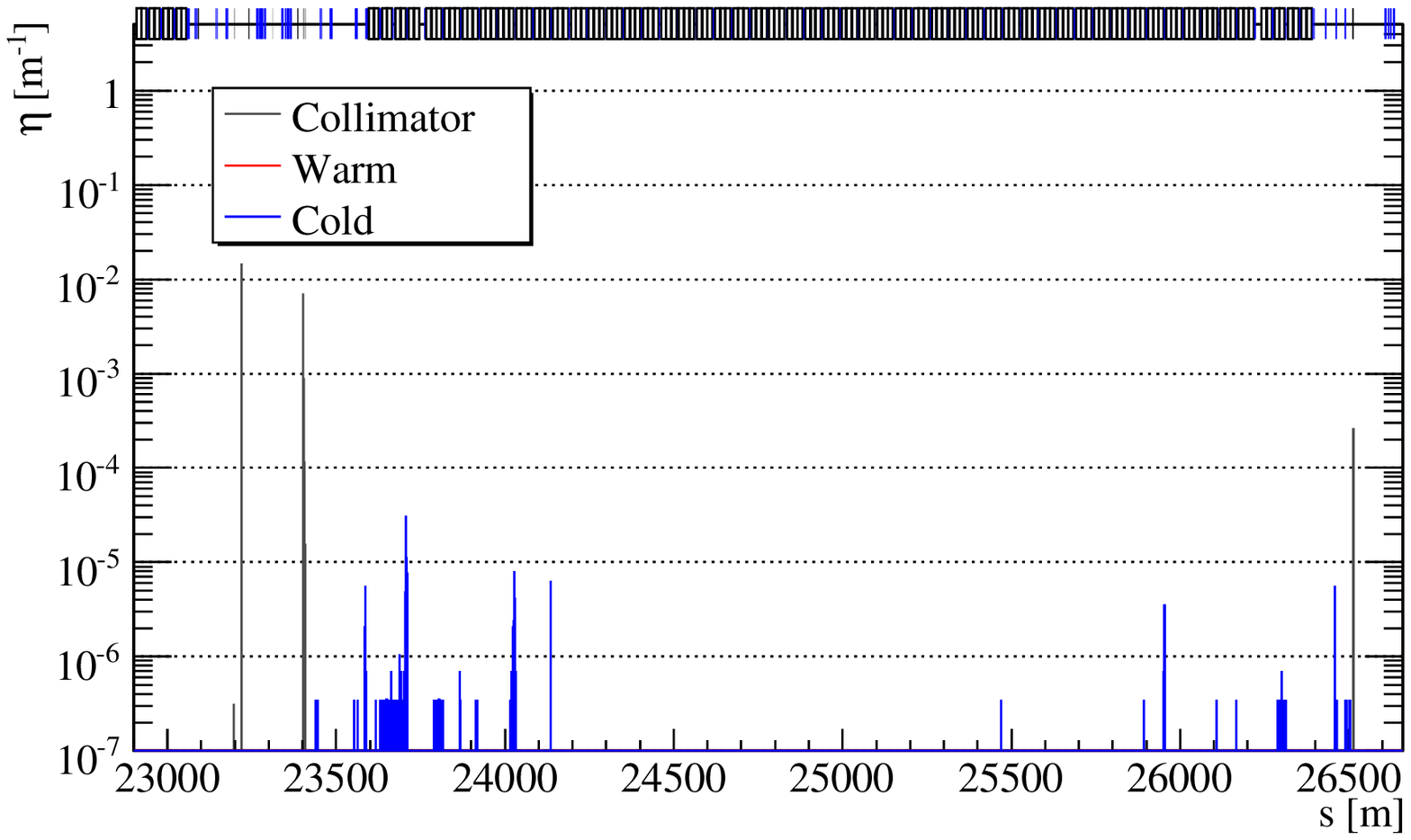}\label{fig:IR8_5p5s_zoom}}
\caption{Simulated beam loss pattern for the IR8 layout with $\mathrm{Cry_1}$ at $5.5~\sigma$. The whole LHC (a), zoom of the arc 81 (b) (1~p~=~3.5$\times10^{-7}~\mathrm{m^{-1}}$).
}
\label{fig:IR8_5p5s}
\end{figure}

\begin{figure}[!t]
\subfloat[]{\includegraphics*[width=90mm]{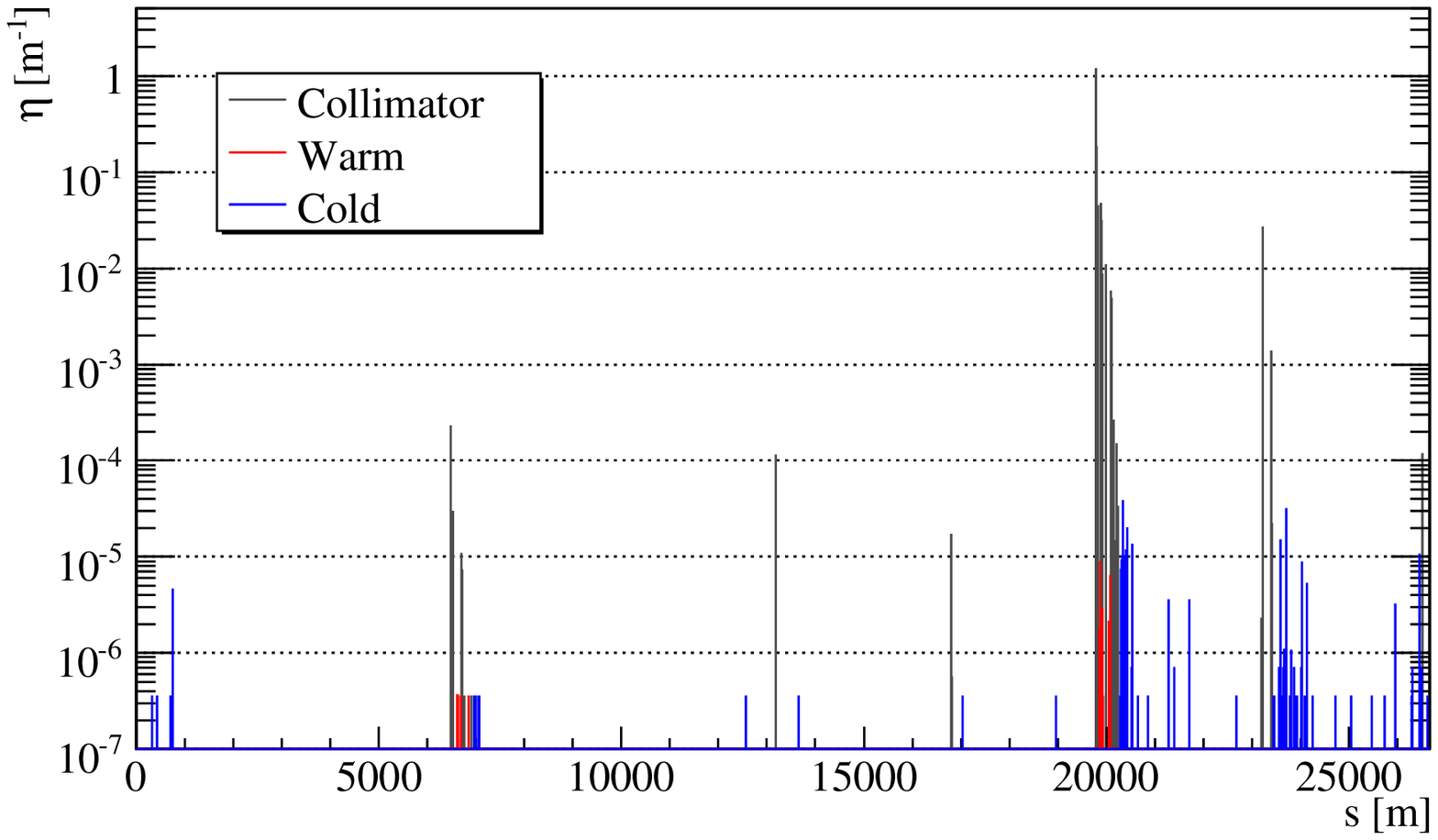}\label{fig:IR8_5p5s_tot_AM}}\vspace*{-\baselineskip}\\
\vspace*{-\baselineskip}
\subfloat[]{\includegraphics*[width=90mm]{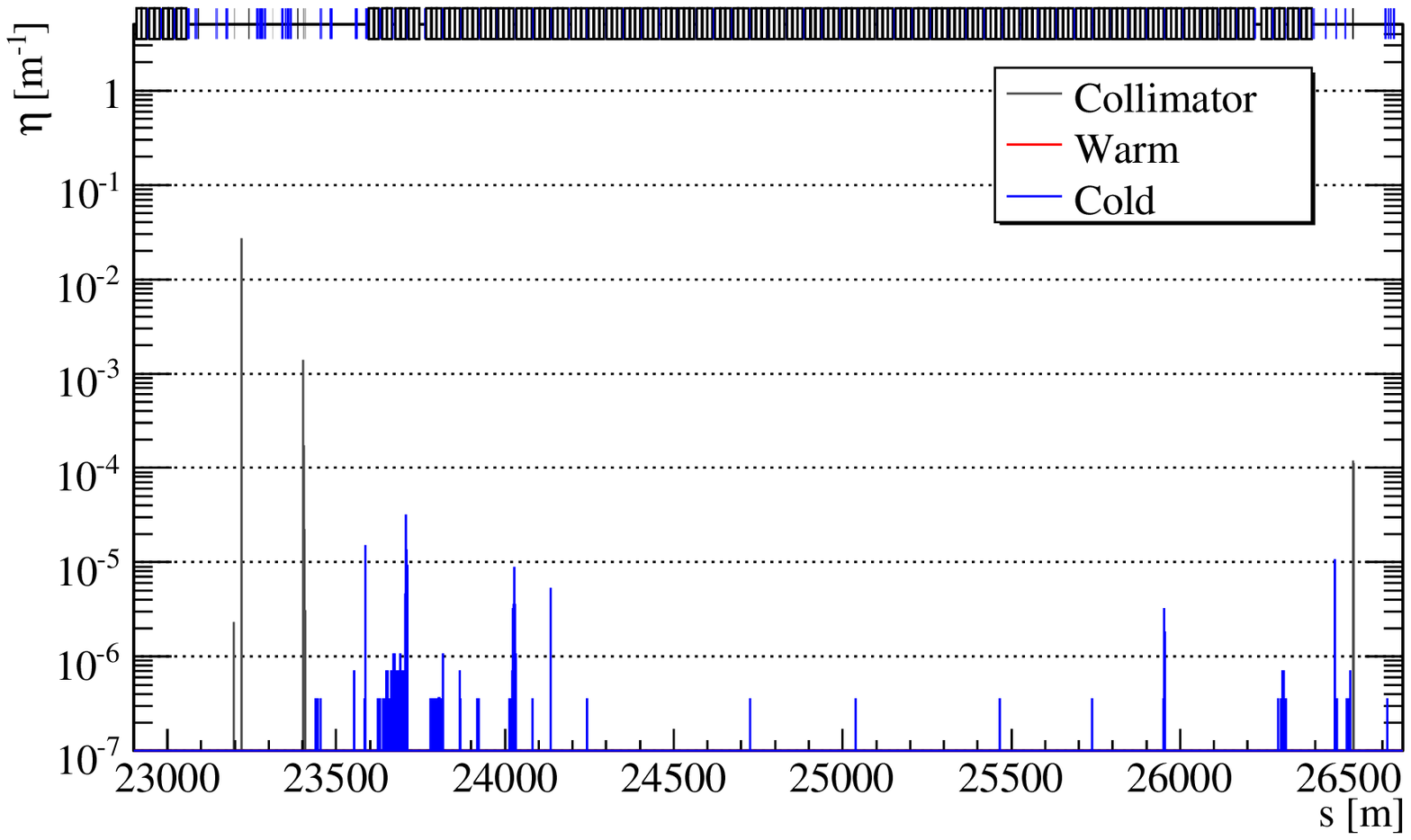}\label{fig:IR8_5p5s_zoom_AM}}
\caption{Simulated beam loss pattern for the IR8 layout with $\mathrm{Cry_1}$ at $5.5~\sigma$ in amorphous orientation. The whole LHC (a), zoom of the arc 81 (b) (1~p~=~3.6$\times10^{-7}~\mathrm{m^{-1}}$).
}
\label{fig:IR8_5p5s_AM}
\end{figure}

Regarding item \textbf{2}, losses around the ring are below safe limits for both layouts with $\mathrm{Cry_1}$ at $5.5~\sigma$. The expected loss pattern for the IR8 layout is shown in Fig.~\ref{fig:IR8_5p5s}. Of course, the IR3 layout is compatible also with this $\mathrm{Cry_1}$ setting, given the loss pattern below safe limits also with $\mathrm{Cry_1}$ at $5.0~\sigma$. The reduced rate of primary protons on the $\mathrm{Cry_1}$ can make possible running with less absorbers (i.e. only one  TCSG and one TCLA locally) for the IR3 layout. Thus, less collimators and relative infrastructure would be needed with a following cost reduction. However, energy deposition simulations are needed for a final assessment of loads on local collimators and magnets, which will be used to define the material budget need to safely absorb the deflected halo by the $\mathrm{Cry_1}$. Simulations were carried out also with $\mathrm{Cry_1}$ in amorphous orientation, i.e. behaving as a 4 mm long scatterer made of silicon, which do not show any anomaly in the loss pattern. The expected loss pattern for the IR8 layout is shown in Fig.~\ref{fig:IR8_5p5s_AM}, as example. 

Finally, item \textbf{3} is fulfilled by the fact that with $\mathrm{Cry_1}$ at $5.5~\sigma$, the load on crystal and absorbers ensures that local losses will be on the shadow of IR7 losses in the case of lifetime drops, without triggering spurious dumps or inducing a magnet quench. Nevertheless, thresholds on allowed local losses before triggering a beam dump request must be carefully evaluated based on energy deposition simulations.

In conclusion, no show-stopper has been identified for $parasitic$ operations with full machine for both layouts with $\mathrm{Cry_1}$ at $5.5~\sigma$. Nevertheless, additional loss clusters with respect standard operations are visible for the layout in IR8 and will need to be addressed with energy deposition simulations, together with the expected increase of background to the ATLAS experiment due to a larger load on the upstream TCTs.



\begin{figure}[t]
\subfloat[]{\includegraphics*[width=90mm]{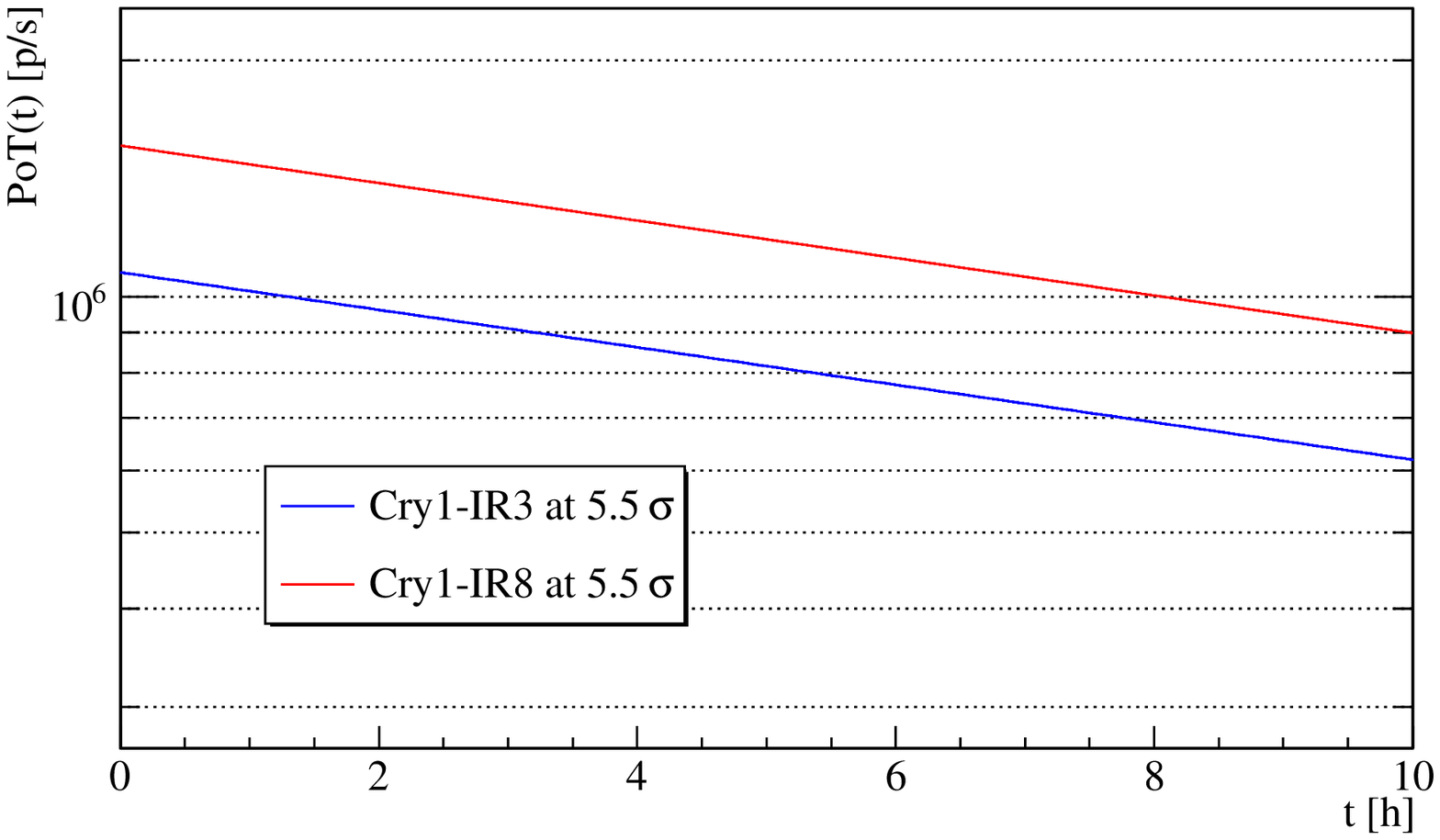}\label{fig:PoT_IR38_inst}}\vspace*{-\baselineskip}\\
\vspace*{-\baselineskip}
\subfloat[]{\includegraphics*[width=90mm]{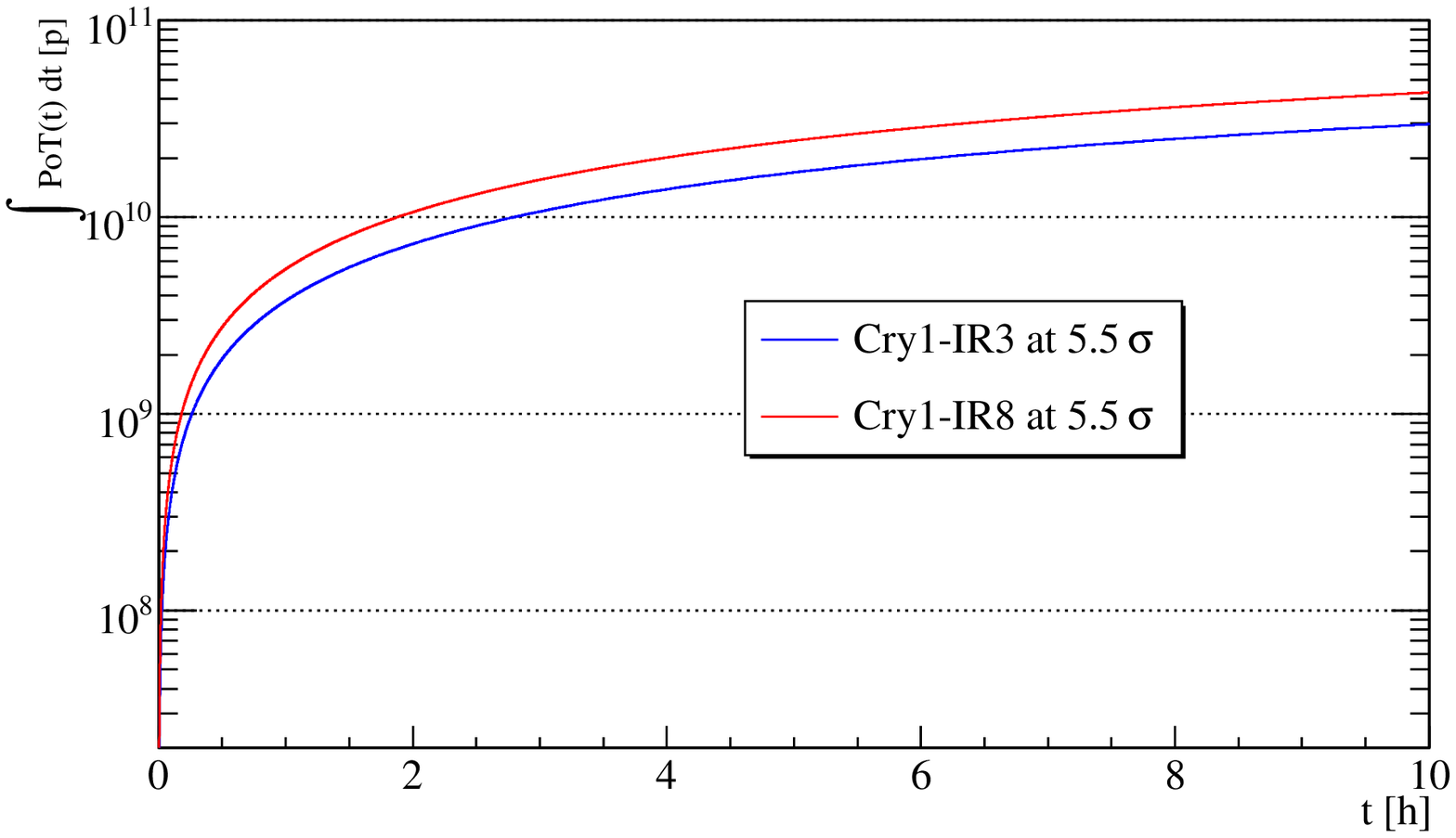}\label{fig:PoT_IR38_int}}
\caption{Expected instantaneous (a) and integrated (b) Proton On Target for both layouts with $\mathrm{Cry_1}$ at $5.5~\sigma$ during a 10~h fill with 200~h beam lifetime.
}
\label{fig:PoT_IR38}
\vspace*{-\baselineskip}
\end{figure}

The achievable instantaneous and integrated $PoT(t)$ during one fill are shown in Fig.~\ref{fig:PoT_IR38}. A summary is reported in Table.~\ref{tab:IR38_fact}. 

\begin{figure}[t]
   \centering
   \includegraphics[width=90mm]{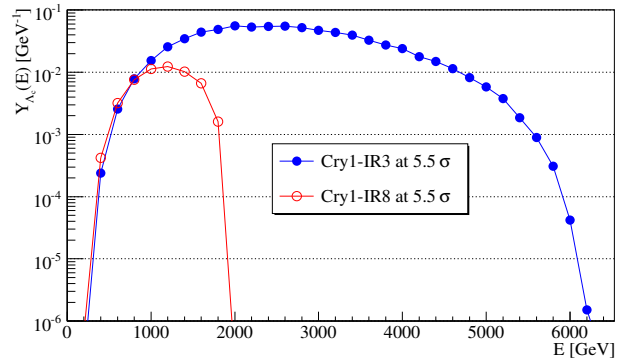}
   \caption{Expected yield of $\mathrm{\Lambda_c}$ emerging from the $\mathrm{Cry_2}$ that acquired the desired precession in 10 h LHC fill with the $\mathrm{Cry_1}$ at $5.5~\sigma$.}
   \label{fig:Lc_yield_5p5s}
\end{figure}

The expected $Y_{\mathrm{\Lambda_c}}(E)$ for $parasitic$ operations of both layouts with full machine and $\mathrm{Cry_1}$ at $5.5~\sigma$ are shown in Fig.~\ref{fig:Lc_yield_5p5s}. About a factor 20 more integrated $\mathrm{\Lambda_c}$ over the available energy spectrum are expected using the IR3 layout with respect to IR8. 

\begin{table}[t]
   \centering
   \caption{Fraction of simulated protons that hit the $\mathrm{Cry_1}$ at $5.5~\sigma$ in both layouts and relative channeling efficiency, together with integrated PoT in a 10 h fill with 200 h beam lifetime.}
   \begin{tabular}{cccc}
       \hline\noalign{\smallskip}
       \textbf{IR} & \textbf{$\frac{N_{\rm{imp}}^{\mathrm{Cry_1}}}{N_{\rm{sim}}}$} & $\varepsilon_{CH}^{\mathrm{Cry_1}}$ & $\int_{10h} PoT(t)dt$ [p]\\
       \noalign{\smallskip}\hline\noalign{\smallskip}
           3         & $1.1\times10^{-2}$            & 0.50     &  $3.0\times10^{10}$  \\ 
           8         & $1.4\times10^{-2}$            & 0.57 &  $4.3\times10^{10}$	 \\ 
       \noalign{\smallskip}\hline
   \end{tabular}
   \label{tab:IR38_fact}
   \vspace*{-\baselineskip}
\end{table}

\subsection{Margins for improvements and further constraints}

Margins for improvements and further constraints are present, but are out of the scope of this paper and are reported as an overview of possible future studies. 

An increased target thickness\footnote{In the approximation of $l_{\rm{t}} \ll c\gamma\tau_{\mathrm{\Lambda_c}}$.} with  and use of germanium crystals can lead to a larger production of $\mathrm{\Lambda_c}$ and channeling efficiency, respectively. Local losses induced by the 5 mm long tungsten target considered in this paper are negligible, but the effect of thicker targets needs to be addressed. Single-pass channeling efficiency can be increased up to about 10\% using germanium with respect to silicon~\cite{de2013highly}.

A factor 2 in bunch intensity is expected if running in the HL-LHC scenario, which is directly translated in a factor 2 larger PoT rate with respect to what considered in previous sections. Moreover, the selective excitation of bunch trains in the vertical plane could lead to larger flux of particles on $\mathrm{Cry_1}$, thus increasing the $PoT(t)$. However, this beam excitation is made through white noise~\cite{ADT} leading to an emittance blow up that will induce a reduction of luminosity in the main experiments. Thus, a compromise between the acceptable loss of luminosity and required increase of $PoT(t)$ will need to be found. 

Additional constraints can come from failure scenarios in HL-LHC. For example, crab-cavities failures can induce large bunch oscillations/rotations~\cite{CCfail}. Thus, a significant fraction of mis-kicked bunches can be intercepted and potentially channeled by $\mathrm{Cry_1}$. Hence, an additional safety margin with respect to TCPs must be computed, which can be larger than the $0.5~\sigma$ considered above.

Constraints can come also from possible damages to the detectors. The LHC collimation system is designed to withstand the failure scenario of beam lifetime drops down 0.2 h for 10 s. With LHC operational settings in 2018, $\mathrm{Cry_1}$ at $5.5~\sigma$ and 200 h of beam lifetime an average PoT rate of about $10^6$ p/s is expected, as shown in Fig.~\ref{fig:PoT_IR38_inst}. Thus, a lifetime of 0.2 h will directly translates in to PoT rate of about $10^9$ p/s. A beam dump request is triggered by LHCb if dangerous events rate are approached. Thus, no damages should be caused but energy simulations are needed to asses this limit in LHCb, to set appropriate margins on sustainable PoT rate during operations and failure scenarios. The layout flexibility in IR3 may allows a detector design capable to withstand higher events rate.

Last but not least, the LHCb experiment is operated at leveled luminosity. The leveling is performed reducing the separation bump in steps. The separation plane in LHCb is vertical. All the components of the proposed layout are also in the vertical plane, and within the separation bump. Thus, dynamic changes of settings (in mm, not in $\sigma$) may be needed during the physics data taking to follow closed-orbit movements. As opposed to IR8, in IR3 everything is frozen once arrived at top energy and the data taking can be carried out in static conditions.

\vspace{-4mm}

\section{Conclusions}

\vspace{-2mm}

Two possible layouts for fixed-target experiments and dipole-moment measurements of short-living baryons at the LHC were presented. Both designs were optimized in order to maximize the number of deliverable PoT, while keeping the losses on superconducting magnets below limits  tolerable for standard LHC operations. A natural choice would be to place this experiment in IR8, to profit of the presence of the LHCb detector. On the other hand, the particular features of this insertion poses several constraints on achievable rate of PoT that cannot be easily overcome. Thus, an alternative layout placed in the momentum cleaning insertion IR3 was studied, showing a significant increase of achievable yield of $\mathrm{\Lambda_c}$. However, a dedicated detector should be built in IR3, for which about 70 m of available longitudinal space have been taken into account in this design.

\begin{acknowledgements}

The authors would like to thank ... for the useful discussions and comments.

D. Mirarchi acknowledge partial support by the High Luminosity LHC project.

A. S. Fomin acknowledge partial support by the National Academy of Sciences of Ukraine (project KPKVK 6541230).

\end{acknowledgements}

\bibliographystyle{spphys}
\bibliography{all_bib}

\end{document}